\documentclass[12pt]{amsart}
\usepackage{amssymb}
\usepackage{eucal}
\usepackage{epsfig}
\usepackage{graphicx}
\usepackage{upref}

\newtheorem{theorem}{Theorem}[section]
\newtheorem{lemma}[theorem]{Lemma}
\newtheorem{cor}[theorem]{Corollary}

\theoremstyle{definition}
\newtheorem{definition}{Definition}[section]

\allowdisplaybreaks
\numberwithin{equation}{section}
\renewcommand{\imath}{\,\mathrm{i}\,}
\begin{document}

\title[Effective integration of the VNSE ]
{Effective integration of the Nonlinear Vector Schr\"odinger Equation }
\author{J N Elgin}
\address{Imperial College, 180 Queen's Gate, London SW7 2BZ}
\email{j.elgin@ic.ac.uk}
\author{V Z Enolski}
\address{Concordia University, 7141 Sherbrooke West, Montreal H4B1R6,
 Montreal, PQ, Canada }
\email{vze@ma.hw.ac.uk }
\author{A R Its}
\address{Department of Mathematical Sciences,
Indiana University - Purdue University Indianapolis, Indianapolis, IN
       46202-3216, USA.  }
\email{itsa@math.iupui.edu }

\begin{abstract}
A comprehensive algebro-geometric integration of the two component
Nonlinear Vector Schr\"odinger equation (Manakov system) is
developed. The allied spectral variety is a trigonal Riemann
surface, which is described explicitly and the solutions of the
equations are given in terms of $\theta$-functions of the surface.
The final formulae are effective in that sense that all entries like
transcendental constants in exponentials, winding vectors etc. are
expressed in terms of prime-form of the curve and well algorithmized
operations on them. That made the result available for direct
calculations in applied problems implementing the Manakov system.
The simplest solutions in Jacobian $\vartheta$-functions are given
as particular case of general formulae and discussed in details.
\end{abstract}

\maketitle
\tableofcontents
\section{Introduction}

The Vector Nonlinear Schr\"odinger equation (VNSE) usefully models
the propagation of a polarized optical beam along an optical fiber.
The vector nature of the dependent variable models the polarization
state of the beam. It is intended in this article to derive and
investigate a general class of periodic and quasi-periodic solutions
of this equation. As the spectral curve for this system is trigonal
-- rather then hyperelliptic as for the scalar case -- existing
formulae for these solutions are rather formal and not tractable for
applications. Here, we use a method first devised by Krichever
\cite{kr77} to effect an explicit integration of the VNSE. This
approach permits us to investigate some special cases where the
formal solutions thus obtained reduced to simpler types expressible
in terms of hyperelliptic or elliptic functions.

In this paper we shall consider the integrable 2 dimensional focusing Vector
Nonlinear Schr\"{o}dinger equation (VNSE)
\begin{eqnarray}
\imath \frac{\partial q_{1}}{\partial t}
+\frac{\partial ^{2}q_{1}}{\partial x^{2}}
+2\left( |q_{1}|^{2}+|q_{2}|^{2}\right) q_{1} &=&0,\label{manakov1} \\
\imath \frac{\partial q_{2}}{\partial t}
+\frac{\partial ^{2}q_{2}}{\partial x^{2}}
+2\left( |q_{1}|^{2}+|q_{2}|^{2}\right) q_{2} &=&0.\label{manakov2}
\end{eqnarray}
It was proven by Manakov \cite{manak74}
that this system is completely integrable and, in consequence,
(\ref{manakov1},\ref{manakov2})
are now known as the {\it Manakov system}.

Manakov's method is based on the Lax
representation
\begin{eqnarray}
\phi_x&=&M\phi,  \label{eq1} \\
\phi_t&=&B\phi,  \label{eq2}
\end{eqnarray}
where
\begin{equation}  \label{mmatrix}
M(z)=\left(
\begin{array}{ccc}
-\imath z & q_1 & q_2 \\
-\bar{q}_1 & \imath z & 0 \\
-\bar{q}_2 & 0 & \imath z
\end{array}
\right)
.\end{equation}
and
\begin{equation}  \label{bmatrix}
B(z)=-\imath \left(
\begin{array}{ccc}
2z^2- |q_1|^2-|q_2|^2 & 2\imath q_1 z -q_{1x} & 2\imath q_2z-q_{2x} \\
-2\imath \bar{q}_1z-\bar{q}_{1x} & -2z^2+|q_1|^2 & \bar{q}_1q_2 \\
-2\imath \bar{q}_2z-\bar{q}_{2x} & q_1\bar{q}_2 & -2z^2+|q_2|^2
\end{array}
\right),
\end{equation}
where bar denotes complex conjugation. The Manakov system can be
represented in the form
\begin{equation}
B_x-M_t=[M,B].  \label{zerocurv}
\end{equation}

The simplest solution \textit{Manakov's soliton} has the form
\begin{eqnarray}\boldsymbol{q}_{sol}(x,t)&=&2\eta
\,\mathrm{sech} (  2\eta(x+4\xi t)  )\label{manakov}\\&\times&
\mathrm{exp}\{  -2\imath\xi x-4\imath (\xi^2-\eta^2)t \}
\boldsymbol{c},\notag\end{eqnarray}
where $\boldsymbol{c}=(c_1,c_2)^T$ is a unit vector, $|c_1|^2+|c_2|^2=1$,
independent of both $x$ and $t$, and $\xi$
and $\eta$ are real constants.

Periodic and quasi-periodic solutions expressed in terms of explicit
$\theta$-functional formulae have been quoted by several authors.
The one component case, i.e. standard nonlinear
Schr\"odinger equation was developed in \cite{its76}, \cite{itskot76},
and \cite{prev85a} (see also monograph  \cite{bbeim94}).
The multi-component case was studied in \cite{kr77,ahh90}.
while the special case of reduction to a dynamical system
with two degree of freedom was studied in \cite{ceek00}.
In recent years attention has been directed to modulation instabilities
of the multi-component equation and searching for homoclinic orbits
\cite{fsw00}, \cite{fmmw00}, \cite{wf00}. Although we are not touching
this interesting and important subject we believe that effective
$\theta$-functional formulae could shed some new light on it.
Indeed, we believe that they will be as useful for studying
homoclinic orbits of the Manakov model as the one component
$\theta$-functional
formulae are for studying the homoclinic orbits of the standard nonlinear
Schr\"odinger
equation (see Sections 4.4 and 4.5 of \cite{bbeim94}).

The article is organized as listed in Contents. The work is a
mixture of analysis and computer algebra implementations using the
Maple code described in \cite{dh01}.

\section{Zero-curvature representation}

Denote by $t_1=x,t_2=t,\ldots,t_n,\ldots$ a set of ``times" and introduce
the set of $3\times3$
matrices $\mathcal{ L}_1(z),\mathcal {L}_2(z),\ldots,\mathcal {L}_n(z),
\ldots $ satisfying  the zero
curvature representation,
\begin{equation}
\frac{\partial}{\partial t_i}\mathcal{L}_j(z) -\frac{\partial}{\partial t_j}
\mathcal{L}_i(z)= [\mathcal{L}_i(z),\mathcal{L}_j(z)], \label{zcr}
\end{equation}
where the  matrices ${\mathcal L}_1$ and ${\mathcal L}_2$
are chosen to satisfy the Lax representation
(\ref{eq1})- (\ref{bmatrix}). More generally, ${\mathcal L}_n(z)$ is expanded
as the $n$-th degree polynomial

\begin{equation}
\mathcal{ L}_n(z)=(2z)^nL_0+(2z)^{n-1}L_1+\ldots+(2z)L_{n-1}+L_n,
\quad n=1,\ldots,
\label{ln}
\end{equation}
having the property
\begin{equation} {\mathcal L}_n^{\dagger}({\bar{z}})=- {\mathcal
    L}_n(z),\label{involution1}
  \end{equation}
where dagger $\dagger$   denotes conjugate transpose:
$\boldsymbol{c}^{\dagger}= \bar{\boldsymbol{c}}^T$.

Here
\begin{eqnarray*}
L_0&=&\left(\begin{array}{cc}-\frac{1}{2}\imath&\boldsymbol{0}^T\\
 \boldsymbol{0}&\frac{1}{2}\imath 1_2\end{array}\right),\quad
L_1=\left(\begin{array}{cc}0&\boldsymbol{q}^T\\
 -\bar{\boldsymbol{q}}&0_2\end{array}\right),\\
L_2&=&\imath\left(\begin{array}{cc}\boldsymbol{q}^T\bar{\boldsymbol{q}}&
\boldsymbol{q}_x^T\\
 \bar{\boldsymbol{q}}_x& -\bar{\boldsymbol{q}}\boldsymbol{q}^T
\end{array}\right)
,\end{eqnarray*}
where
\[ \boldsymbol{q}=(q_1,q_2)^T,\]
while, for $k>2$ introduce the following ansatz
 \begin{equation}L_{k}=\left(
\begin{array} {cc} \alpha_k&\boldsymbol{\beta}_{k-1}^T\\
\boldsymbol{\gamma}_{k-1}&   {\mathcal A}_k
\end{array}\right) +L_k^{(0)}.\label{li}
\end{equation}
In equation (\ref{li})  $\mathcal A$ denotes a $2\times2$-matrix and
$L_k^{(0)}$ denotes a constant matrix of the form
\begin{equation}
L_k^{(0)}=\left(\begin{array}{ccc}c_{1,1}^k&0&0\\
                                            0&c_{2,2}^k&c_{2,3}^k\\
                                            0&c_{3,2}^k&c_{3,3}^k\end{array}\right)
\label{Cmatrix}
\end{equation}
with arbitrary entries $c_{p,q}^k$. In this article we set $L^{(0)}_k=0$.

The following theorem is valid

\begin{theorem} The entries to the matrix (\ref{li}) in
the zero-curvature representation
are defined as follows
\begin{itemize}
\item the vectors $\boldsymbol{\beta}_n$ and  $\boldsymbol{\gamma}_n$
are given by the equations
\[
\begin{aligned}
\boldsymbol{\beta}_n&
=\left(\imath D\right)^n\boldsymbol{q},\quad
\boldsymbol{\gamma}_n = -\bar{\boldsymbol{\beta}}_n,
\end{aligned}
\]
where $D$ acts on a vector $\boldsymbol{f}(x)$  as
\begin{equation}
D\boldsymbol{f}(x) =\frac{\partial}{\partial x}\boldsymbol{f}(x)
+\int\limits_.^x
\left\{ \boldsymbol{q}(x')^{\dagger}, \boldsymbol{f}(x')
\right\}_A dx'\boldsymbol{q}(x),
\end{equation}
where $\{\cdot,\cdot\}_A$
denotes the matrix which is the anti-hermitian part of
anticommutator, so that
\begin{equation}\{\boldsymbol{a}^{\dagger},\boldsymbol{b}\}_A
=\left(\boldsymbol{a}^{\dagger}\boldsymbol{b}-
\boldsymbol{b}^{\dagger}\boldsymbol{a}\right)1_2
    + \boldsymbol{b}\boldsymbol{a}^{\dagger}
-\boldsymbol{a}\boldsymbol{b}^{\dagger}.\label{abracket}
\end{equation}
 Therefore, the flows are defined as

\begin{align}
\boldsymbol{q}_n
\equiv \frac{\partial}{\partial t_n} \boldsymbol{q}
=\imath \left(\imath D\right)^n\boldsymbol{q}. \label{qn}
\end{align}

\item
The $(1,1)$ element of the matrix $L_{k+2}$
is defined recursively as follows

\begin{equation}
\alpha_{k+2}=-\imath \sum_{j=0}^{k}\boldsymbol{\gamma}^T_{k-j}
\boldsymbol{\beta}_j
-\imath\sum_{j=0}^{k-2}\alpha_{k-j}\alpha_{j+2},\label{alpha}
\end{equation}
with $$\alpha_0=-\frac{\imath}{2}, \quad \alpha_1=0,\quad
\alpha_2=\imath \boldsymbol{q}^{T}\bar{\boldsymbol{q}},$$
while the associated right lower $2\times2$ minor
${\mathcal A}_{k+2}$ is given recursively as
\begin{equation}
{\mathcal A}_{k+2}=\imath
\sum_{j=0}^{k}\boldsymbol{\gamma}_{k-j}\boldsymbol{\beta}_j^T
+\imath\sum_{j=0}^{k-2}\alpha_{k-j}{\mathcal A}_{j+2}\label{alpha1}
\end{equation}
with  $${\mathcal A}_0=\frac{\imath}{2} 1_2,
 \quad {\mathcal A}_1=0,
\quad {\mathcal A}_2=-\imath \bar{\boldsymbol{q}}\boldsymbol{q}^T.$$
\end{itemize}
In each case, contributions from the second sum appear only for $k\geq
2$.
\end{theorem}
\begin{proof}
 The proof of these results follows from the substitution
of ansatz (\ref{ln}), (\ref{li}) into (\ref{zcr}) with
$i = 1$ and $j = n$ and solving the equation recursively.
Also, one has to take into account that
$\frac{\partial}{\partial
t_i} \mathrm{Tr}({\mathcal L}_n^2)=0$.
\end{proof}

\section{The spectral curve}

The spectral curve is fixed by defining the stationary flow as follows:
let the system depend only on times $t_1,\ldots,t_{n-1}$. Then the
zero curvature representation (\ref{zcr}) written for $\mathcal{L}_1$ and
$\mathcal{L}_n$ has the form
\begin{equation}
\frac{\partial}{\partial x}\mathcal{L}_{n}(z)[\mathcal{L}_1(z),\mathcal{L}_{n}(z)]. \label{zcr1}
\end{equation}
This relation suggests we consider the polynomial equation
\begin{equation}
 f(z,w)=0,\qquad  f(z,w)= \mathrm{det}( \mathcal{L}_{n}(z) -w
1_3) \label{curve}.
\end{equation}
We shall call the polynomial equation
\begin{equation}X:=\{(z,w)\vert  f(z,w)=0  \}. \end{equation}
the spectral curve.
Evidently
coefficients of monomials $z^kw^l$ of the polynomial $f(z,w)$ are
constants of motion. In what follows we shall consider the Riemann
surface of the curve $X$, which we shall denote by the same letter.

To proceed we recall that any rational function of its arguments,
$\phi(z,w)$ is called a
function on the curve $f(z,w)=0$. The order of the function $\phi(z,w)$
on the curve $X$ is the number $N$ of
common zeros $(z_1,w_1),   \ldots,   (z_N,w_N)$
of equations $f(z,w)=0$ and $\phi(z,w)=0$. The curve is hyperelliptic if it
admits a function of the second order, it is trigonal if it admits a function
of third order etc.

In the case considered, the spectral curve
can be written in the explicit form as
\begin{eqnarray}
&&X=\{(z,w)\vert  f(z,w)=0\},\notag \\
&&f(z,w)=(w+\frac{\imath}{2} (2z)^n)
(w-\frac{\imath}{2} (2z)^n)^2\cr
&&+(w-\frac{\imath}{2} (2z)^n)\sum_{j=n}^{2n-1} \lambda_j(2z)^{2n-j-1}
+\sum_{j=0}^{n-2}\mu_{n-2-j}(2z)^j,\label{trigonal}
\end{eqnarray}
where $2n-1$ parameters $\lambda_i$ $i=n,\ldots,n-1$
and $\mu_j$, $j=0,\ldots,n-2$ are constants of motion and can be
taken arbitrary, but satisfying conditions given below in (\ref{lmbar}).
The coordinate $z$ of the curve is a function of the
third order and therefore the curve is trigonal.

The parameters
$\lambda_j$ of the curve $X$ can be computed
in terms of $\boldsymbol{q}$ as follows:
\begin{eqnarray}
\lambda_j&=&-\frac{1}{2}\sum_{k=j-n}^{n-1}
\boldsymbol{\gamma}^T_k\boldsymbol{\beta}_{j-k-1}\notag\\
&-&\frac{1}{2}\sum_{k=j-n-1}^{n-2} \alpha_{k+2}\alpha_{j-k-1},\quad
j=n,\ldots, 2n-1.\label{recursion}
 \end{eqnarray}
In particular,
\begin{eqnarray}
\lambda_n&=&\alpha_0\alpha_{n+1},\\
\lambda_{n+1}&=&\alpha_0\alpha_{n+2}+\frac{1}{2}\boldsymbol{\gamma}^T_n\boldsymbol{\beta}_0
                           +\frac{1}{2}\boldsymbol{\gamma}^T_0\boldsymbol{\beta}_n.
\end{eqnarray}
The structure of the second term in (\ref{trigonal}) has been obtained
analytically, including the stated expressions for $\lambda_j$.
By contrast,information concerning the final term has been obtained
using Maple, which gives the polynomial structure of degree $n-2$ indicated.

It follows from (\ref{involution1}) that the curve $X$
admits the anti-involution property
\begin{equation} \sigma: X\longrightarrow X,\quad \text{where}\quad
\sigma:(z,w)\rightarrow (\bar{z},-\bar{w})
\label{taudefinition}
\end{equation}
That implies in accordance with explicit formula for $\lambda_i$
\begin{align}\begin{split}
\bar{\lambda}_i&=\lambda_i,\quad i=n,\ldots, 2n-1,\\
\bar{\mu}_j&=-\mu_j,\quad j=0,\ldots, n-2.\end{split}\label{lmbar}
\end{align}
Therefore we have
\begin{equation}\sigma\circ
  f(z,w)=f(\bar{z},-\bar{w})=-\overline{f(z,w)},
\end{equation}
what means that the curve $X$ has required anti-involution property.

Let us clarify now the question on the genus $g$  of the curve $X$.
\begin{lemma}\label{genuslemma} Let $L_n^{(0)}=0$ and
the curve $X$ is given by the equation (\ref{trigonal})
with parameters $\lambda_i, \mu_j$ in general position. Then the genus
of $X$ is given by the formula
\begin{equation} g=2n-3.\label{genusX}\end{equation}
\end{lemma}

\begin{proof} Write equation (\ref{trigonal}) in the form
\begin{align}
&f(z,w)=(w+\frac{\imath}{2} (2z)^n)
(w-\frac{\imath}{2} (2z)^n)^2\notag\\
&+(w-\frac{\imath}{2} (2z)^n)P_{n-1}(z)
+P_{n-2}(z)=0,\label{trigonal1}
\end{align}
where $P_{n-1}(z)$ and $P_{n-2}(z)$ are polynomials of degrees $n-1$
and $n-2$ correspondingly. The discriminant of (\ref{trigonal1}) be of
the form
\begin{align*}
&\mathrm{Discriminant}(X)=\mathrm{Resultant} \left(f(z,w),
\frac{\partial}{\partial w} f(z,w),
w  \right)\\
&=256\imath
P_{n-2}(z)z^{3n}+16P_{n-1}(z)^2z^{2n}+27P_{n-2}(z)^2\\&+72
\imath P_{n-2}(z)P_{n-1}(z)z^n+16P_{n-1}(z)^2z^{2n}+4P_{n-1}(z)^3.
\end{align*}
The degree in $z$ of the $\mathrm{Discriminant}(X)$ be $4n-2$ because
coefficient of the leading power be
\begin{equation} \lambda_n^2-4\imath \mu_0 \neq 0,\quad n=2,3,\ldots
 \label{leadpow}
 \end{equation}
for $\lambda_i, \mu_j$ in general position. Moreover for general
values of parameters the $\mathrm{Discriminant}(X)$ has no multiple
roots and all zeros are simple branch points of the curve $X$. Beside
of that we remark that the curve $X$ has no branch points at
infinities, $\infty_1,\infty_2,\infty_3$
Therefore the curve $X$ has $4n-2$ simple branch points altogether,
which we will denote $e_1$, $e_2$, ..., $e_{4n-2}$ .
The application of the Riemann-Hurwitz formula
\begin{equation} g=\frac{B}{2}-N+1,   \end{equation}
where $B$ is total branch number, being equal in the case $4n-2$ and
$N$ is the number of sheets of the cover over Riemann sphere, which is
3 in the case, completes the proof.
\end{proof}

We remark that our formula for genus (\ref{genusX}) is addressed to
the concrete curve which is fixed for our analysis. The inclusion of
constant matrices $L_k^{(0)}$ can increase the genus. The discrepancy
of our formulae with results of \cite{ahh90} and \cite{wright99} is
due to the fact that in there an estimate of upper bound for genus was
given for more general curve then our be.

Introduce further the Riemann surface of the curve $X$. To do that we
define local coordinate  $\xi(P)$ of a point $P=(x,y)\in X$ in
vicinity of another point $P=(z,w)\in X$ as follows
\begin{equation}
x=\begin{cases}z+ \xi&\text{if}\; P=(z,w)\; \text{is regular
    point},\\
a+\xi^2  &\text{if}\; P=(a,w(a))\; \text{is branch point},\\
\frac{1}{\xi}&\text{if}\; P=(\infty,\infty)\; \text{is regular
    point at infinity}.\\
\end{cases}
\end{equation}
To comment this definition we remark that for general values of
parameters $\lambda_i$ and $\mu_i$ the curve has only simple branch
points with ramification number one, what leads to the structure of
the second line of the definition. The curve $X$ has 3-sheeted
structure with regular points at infinities,
$\infty_1,\infty_2$ and $\infty_3$ where the coordinate of the curve
behave as follows

\begin{align}\begin{split}
z=\frac{1}{\xi},\quad&
w=-\frac{2^{n-1}\imath}{\xi^n}-\frac{\imath}{2}\lambda_n\xi+O(\xi^2)\quad\text{on
the first sheet},\\
z=\frac{1}{\xi},\quad&
w=\frac{2^{n-1}\imath}{\xi^n}+\frac{\imath}{4}(\lambda_n+\sqrt{\lambda_n^2-4\imath\mu_0})
\xi+O(\xi^2)\quad\text{on the second sheet},\\
z=\frac{1}{\xi},\quad&
w=\frac{2^{n-1}\imath}{\xi^n}+\frac{\imath}{4}(\lambda_n-\sqrt{\lambda_n^2-4\imath\mu_0})
\xi+O(\xi^2)\quad\text{on the third sheet}.\end{split}\label{sheets}
\end{align}

We shall also assume that the branch points are all complex,
form the conjugated pairs, i.e.
\begin{equation}\label{branchpoints}
\bar{e}_{2k-1} = e_{2k}, \quad \mbox{Im}\,e_{2k-1} <0,
\end{equation}
and Re $e_{2k-1} <$ Re $e_{2k+1}$.

We are in position now to introduce a suitable homology
basis on the Riemann surface of the curve
(\ref{trigonal}). A canonical basis of cycles  $\mathfrak{a}_i$
and $\mathfrak{b}_i$
respecting intersection property $\mathfrak{a}_i\circ \mathfrak{a}_j =0$,
$\mathfrak{b}_i
\circ \mathfrak{b}_j =0$,
$\mathfrak{a}_i\circ \mathfrak{b}_j =        -\mathfrak{b}_i\circ
\mathfrak{a}_j    =\delta_{ij}$ which also respect the involution
property
\begin{align}
\sigma(\mathfrak{a}_j)&=-\mathfrak{a}_j,\label{sigmacycles}\\
\sigma(\mathfrak{b}_j)&=\mathfrak{b}_j-2\mathfrak{a}_j-\sum_{k\neq j}
 \mathfrak{a}_k.\notag
\end{align}

The homology basis for the case $g=3$ is shown in figure
1; here, the solid, dashed and dash-dotted lines connecting points $e_1$
to $e_2$ etc. are cuts connecting the first to second, second to third
and third to first sheets respectively.  See caption for further
comments. The homology  basis for higher genera can be plotted analogously.

\begin{figure}[ht]
\includegraphics[width=0.8\textwidth]
{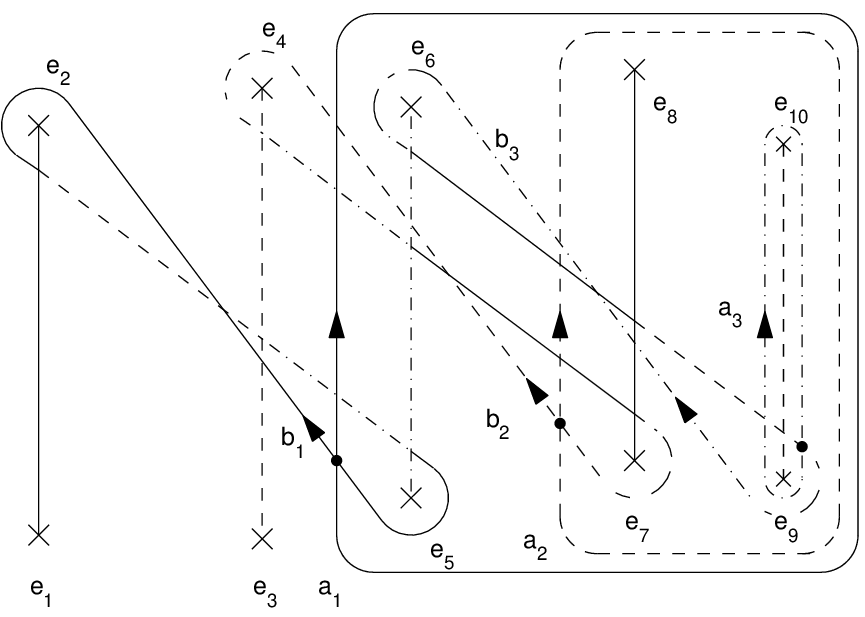}
\caption{Basis of cycles of the curve $X$ of  genus $3$.
The solid, dashed and dash-dotted
lines denote paths on the first, second and third sheets respectively.
Correspondingly
the solid to dashed line, dashed to dot-dashed line, and dot-dashed to solid
lines illustrate trajectories passing through these cuts.
The cuts are similarly encoded for clarity.
}
\label{fig3}
\end{figure}

\begin{figure}[ht]
\includegraphics[width=0.8\textwidth]
{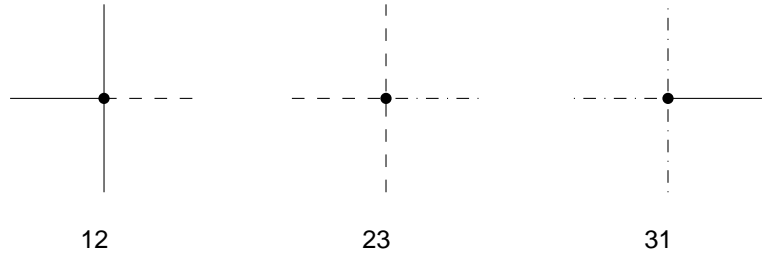}
\caption{Contours passing from sheet 1 to sheet 2, from sheet 2 to sheet 3
and from  sheet 3 to sheet 1}
\label{fig2}
\end{figure}

\section{Differentials and Integrals}

\subsection{Holomorphic differentials and integrals}
Let $X$ be algebraic curve (\ref{trigonal}) of genus $g$ and let
$d\boldsymbol{u}(Q)=(du_1(Q),\ldots,
du_g(Q))^T $ be the set of canonical holomorphic
differentials, which are given at $n>2$ explicitly as

\begin{align}\begin{split}
du_j(Q)&= \frac{\imath z^{j-1}}{\frac{\partial}{\partial w} f(z,w) }dz,\quad
j=1,\ldots,n-2,\\ du_j(Q)&=\frac{z^{2n-3-j}
\left(w-\frac{\imath}{2}(2z)^n\right)
  }{\frac{\partial}{\partial w} f(z,w)}\,dz,\quad j=n-1,\ldots,2n-3,
\end{split}
\label{holdiff}
\end{align}
where $f(z,w)$ be the polynomial defining the curve (\ref{trigonal}).
At $n=2$ the curve is elliptic; this case is studied in detail in
Section 6.

If $du = \phi(z,w)dz$ is an Abelian differential, then the action of
the involution $\sigma$, $\sigma^{\ast}$ say,  is defined by the
relation
$$
 \sigma^{\ast}du=\phi(\bar{z},-\bar{w})
\overline{dz}.
$$
For the differentials $du_{j}$ we have that
\begin{equation} \sigma^{\ast}du_{j}= -
\overline{du_{j}}.\label{invprop1}
\end{equation}
Introduce the matrix of $\mathfrak{a}$-periods,
\[ A=\left(\oint\limits_{\mathfrak{a}_j}du_k
\right)_{j,k=1,\ldots,g} .
 \]
From the properties (\ref{sigmacycles}) and (\ref{invprop1}) of
$\sigma^{\ast}$ it
follows that all elements of the matrix $A$ are real.
Normalized form of the above
differential is introduced as
\begin{equation} dv_j=\sum_{l=1}^g C_{jl}du_l,
\label{holnorm} \end{equation}
where the matrix $C=A^{-1}$. Evidently,
\begin{equation} \overline{dv_j}=-\sigma^{\ast}dv_j.
\label{vcon}
\end{equation}

Now introduce $\tau$-matrix, as a matrix of $\mathfrak{b}$-periods of
normalized differentials,
\begin{equation}
\tau=\left(\oint\limits_{\mathfrak{b}_j}dv_k
\right)_{j,k=1,\ldots,g} .\label{tau}
\end{equation}
Then using (\ref{sigmacycles})  and (\ref{vcon}) we find
\begin{align*}
\overline{\tau_{jk}}&=\oint\limits_{\mathfrak{b}_j}
\overline{dv}_k = -\oint\limits_{\mathfrak{b}_j}
\sigma^{\ast}dv_k = -\oint\limits_{\sigma(\mathfrak{b}_j)}dv_k\\
&=-\tau_{jk}+2\delta_{jk}+\sum_{l\neq j}\delta_{lk}.
\end{align*}
In other words we have
\begin{equation}\label{tauinv}
\overline{\tau}=-\tau+\tau_0,
\end{equation}
where all diagonal elements of $\tau_0$ are 2 and all off-diagonal elements
are 1.

\subsection{Meromorphic differentials and integrals}
Our construction is based on the existence of certain Abelian
integrals
$\Omega_1(Q)$ and $\Omega_2(Q)$ of the second kind,  and  similar
integrals
$h_2(Q)$ and $h_3(Q)$ of the third kind. We shall define these integrals
as follows:

\begin{definition}
Define normalized Abelian integrals $\Omega_1(Q)$ and $\Omega_2(Q)$ of
the second kind
\begin{equation}
\oint_{{\mathfrak a}_j}d\Omega_1=0,\quad \oint_{{\mathfrak a}_j}d\Omega_2=0,
\quad j=1,\ldots,g\label{normalsecond}
,\end{equation}
where   $d\Omega_1(Q)$, $d\Omega_2(Q)$,
are the second kind Abelian differentials in such a way that integrals
$\Omega_1(Q)$, and  $\Omega_2(Q)$,
have poles at the infinities  $\infty_1$, $\infty_2$, $\infty_3$
in the vicinity of which the following expansions
are valid
\begin{align}
\Omega_1(Q)&=\begin{cases}
-\imath z + \frac12(E_1+E_2)+O\left(\frac{1}{z}\right)&\text{at}\quad
Q\longrightarrow \;\infty_1,\\
\imath z + \frac12(E_2-E_1)+O\left(\frac{1}{z}\right) &  \text{at}\quad
Q\longrightarrow\;\infty_2\\
\imath z -\frac12(E_2-E_1)+O\left(\frac{1}{z}\right) &  \text{at}\quad
Q\longrightarrow\;\infty_3;
 \end{cases}\label{omega1}
\end{align}
and
\begin{align}
\Omega_2(Q)&=\begin{cases}
-2\imath z^2 - \frac12(N_1+N_2) & \text{at}\quad
Q\longrightarrow\;\infty_1,\\
2\imath z^2 - \frac12(N_2-N_1)& \text{at}\quad
Q\longrightarrow\;\infty_2,\\
2\imath z^2 +\frac12(N_2-N_1) & \text{at}\quad
Q\longrightarrow\;\infty_3,
 \end{cases}\label{omega2}\end{align}
where
  $E_1,E_2$ and $N_1,N_2$ are certain constants.
\end{definition}

We now compute $\mathfrak b$-periods for the second kind differentials.
Let  $P$ be a point in the vicinity of $\infty_i$ and $\xi$ be
the local coordinate. Then the expansion of the vector of normalized
holomorphic integrals reads
\begin{equation}
\left.\int\limits_Q^Pd\boldsymbol{v}\right|_{P\to\infty_i}
=\boldsymbol{U}_i+\boldsymbol{V}^{(i)}\xi+
\boldsymbol{W}^{(i)}\xi^2+\boldsymbol{Z}^{(i)}\xi^3
+\ldots,\quad i=1,2,3,
\label{hexpan}\end{equation}
where $\boldsymbol{U}_i$ are constant vectors depending on the
initial point $Q$  and $\infty_i$,
\[ \boldsymbol{U}_i=\int\limits_Q^{\infty_i}d\boldsymbol{v}.   \]

    It follows from the Bilinear Riemann Relation written for the
differentials $d\Omega_1, d v_j$ and  $d\Omega_2, d v_j$, that
\begin{align}\label{wind}
\boldsymbol{V}&=\imath \boldsymbol{V}^{(1)}-\imath \boldsymbol{V}^{(2)}
-\imath \boldsymbol{V}^{(3)},\\
\boldsymbol{W}&=4\imath \boldsymbol{W}^{(1)}-4\imath\boldsymbol{W}^{(2)}
-4\imath \boldsymbol{W}^{(3)},
\end{align}
where the vectors $\boldsymbol{V}$ and $\boldsymbol{W}$ are defined as
\begin{align}
V_j&=\frac{1}{2\imath \pi}\oint\limits_{\mathfrak{b}_j}d\Omega_1(Q),\quad
W_j=\frac{1}{2\imath \pi}\oint\limits_{\mathfrak{b}_j}d
\Omega_2(Q),\quad j=1,\ldots,g.\label{windings}
\end{align}

We shall refer below to the winding vectors $\boldsymbol{V}$ and
$\boldsymbol{W}$ as the main winding vectors while
the vectors $\boldsymbol{V}^{(i)}$ and
$\boldsymbol{W}^{(i)}$, $i=1,2,3$ we shall call auxiliary winding
vectors.

It is easy to see that the differentials $d\Omega_1$ and $d\Omega_2$
satisfy the same symmetry property (\ref{vcon}) as the differentials
$dv_{k}$. Hence, similar to the derivation of (\ref{tauinv}),
we arrive to the relations
\begin{equation}\label{VWcon}
\overline{V_{j}} = V_{j}, \quad \overline{W_{j}} = W_{j},
..., j = 1, ..., g.
\end{equation}
Moreover, we claim that
\begin{equation}\label{ENcon}
\overline{E_{k}} = - E_{k}, \quad \overline{N_{k}} = - N_{k},
..., k = 1, 2.
\end{equation}

To prove the symmetry properties (\ref{ENcon}) let us notice
that the constants $E_{1,2}$ and $N_{1,2}$ can be determined
via the asymptotic relations,
\begin{equation}\label{E1}
\int_{Q_{2}}^{Q_{1}}d\Omega_{1}
= -2\imath z + E_{1} + O\left(\frac{1}{z}\right),
\quad z \to \infty,
\end{equation}
\begin{equation}\label{E2}
\int_{Q_{3}}^{Q_{1}}d\Omega_{1}
= -2\imath z + E_{2} + O\left(\frac{1}{z}\right),
\quad z \to \infty,
\end{equation}
\begin{equation}\label{N1}
\int_{Q_{2}}^{Q_{1}}d\Omega_{2}
= -4\imath z^2 - N_{1} + O\left(\frac{1}{z}\right),
\quad z \to \infty,
\end{equation}
\begin{equation}\label{N2}
\int_{Q_{3}}^{Q_{1}}d\Omega_{2}
= -4\imath z^2 - N_{2} + O\left(\frac{1}{z}\right),
\quad z \to \infty,
\end{equation}
where the point $Q_{j}$ belongs to the $j$-th sheet of the Riemann surface
$X$,
$$
\quad \pi(Q_{1}) = \pi(Q_{2}) = \pi(Q_{3}) = z,
$$
and $\pi: X \to {\Bbb C}$ is the canonical covering map.
We assume that $z$ is a large real positive number and that the contours
of integration in the integrals $\int_{Q_{2}}^{Q_{1}}$ and
$\int_{Q_{3}}^{Q_{1}}$ do not intersect the
basic cycles and do not pass through the points
$\infty_{j}$. We shall denote these contours  $\mathfrak{l}_{2}$
and
 $\mathfrak{l}_{3}$, respectively. The involution
$\sigma$ acts on the contours $\mathfrak{l}_{j}$ as follows,
\begin{equation}\label{involl2}
\sigma(\mathfrak{l}_{2}) = \mathfrak{l}_{2} + \sum_{\nu
=1}^{g}n^{(2)}_{\nu}\mathfrak{a}_{\nu} + m^{(2)}\alpha_{1} +
n^{(2)}\alpha_{2} +
k^{(2)}\alpha_{3},
\end{equation}
\begin{equation}\label{involl3}
\sigma(\mathfrak{l}_{3}) = \mathfrak{l}_{3} + \sum_{\nu
=1}^{g}n^{(3)}_{\nu}\mathfrak{a}_{\nu} + m^{(3)}\alpha_{1} +
n^{(3)}\alpha_{3} + k^{(3)}\alpha_{2},
\end{equation}
where $\alpha_{j}$ denotes a positively oriented circle around
the point $\infty_{j}$ and   $m^{(2,3)}$,  $n^{(2,3)}$, $k^{(2,3)}$, and
$n_{\nu}^{(2,3)}$ are integers. It should be also noticed, although we
won't use it right now, that the integers  $m^{(2,3)}$ and  $n^{(2,3)}$
have the property{\footnote{In the $g =3$ example featured in the figure 1,
the contour  $\mathfrak{l}_{2}$ starts at the point $Q_{2}$ on the second
sheet,
goes below  the branch points to the branch point $e_{1}$, passes to the
first
sheet and goes below  the branch points to the point $Q_{1}$.
The contour $\mathfrak{l}_{3}$ starts at the point $Q_{3}$ on the
third sheet, goes below the branch points to the branch point $e_{3}$, passes
to the second sheet, goes to the branch point $e_{1}$, passes to the
first sheet and goes below  the branch points to the point $Q_{1}$.
The specifications of the integers $m^{(2,3)}$,  $n^{(2,3)}$, and $k^{(2,3)}$
in equations (\ref{involl2}) and (\ref{involl3}) are:
$$
n_{1}^{(2)} = 1, \quad n_{2}^{(2)} = n_{3}^{(2)} = 0;
\quad m^{(2)} = k^{(2)}=0,\quad n^{(2)} = -1,
$$
$$
n_{1}^{(3)} = n_{2}^{(3)} = 1, \quad n_{3}^{(3)} = 0;
\quad m^{(3)} = 0, \quad n^{(3)} = k^{(3)} = -1.
$$}},
\begin{equation}\label{odddif}
m^{(j)} - n^{(j)} = \mbox{odd number}, \quad j = 2, 3.
\end{equation}
From  (\ref{involl2}) and (\ref{involl3}) we obtain that
$$
\overline{\int_{Q_{j}}^{Q_{1}}d\Omega_{k}}
=\oint_{\mathfrak{l}_{j}}\overline{d\Omega_{k}}=
-\oint_{\mathfrak{l}_{j}}\sigma^{*}d\Omega_{k} =
-\oint_{\sigma(\mathfrak{l}_{j})}d\Omega_{k}
$$
$$
= -\oint_{\mathfrak{l}_{j}}d\Omega_{k} =
-\int_{Q_{j}}^{Q_{1}}d\Omega_{k}, \quad j =2, 3, \quad k =1, 2,
$$
and (\ref{ENcon}) follows in virtue of the asymptotic (\ref{E1}) -
(\ref{N2}).
\begin{definition}
Define normalized Abelian integrals
$h_{2}(Q)$ and $h_{3}(Q)$  of the third kind
\begin{equation}
\oint_{{\mathfrak a}_j}d h_2=0,\quad \oint_{{\mathfrak a}_j}d h_3=0,
 \quad j=1,\ldots,g.\label{normalthird}
\end{equation}

The integral $h_{2}(Q)$ has logarithmic singularities only and only at
infinities on the first and second sheets,
$\infty_{1,2}$ where it behaves locally as

\begin{align}
h_2(Q)&=\begin{cases}
\mathrm{ln}\,z - \mathrm{ln}\delta_2+o(1) &\text{at}\quad
   Q\longrightarrow \infty_2,\\
-\mathrm{ln}\,z + \mathrm{ln}\delta_2+o(1) &\text{at}\quad
   Q\longrightarrow \infty_1.
\end{cases}\label{h2}
\end{align}

The integral $h_{3}(Q)$ has logarithmic singularities
only and only at infinities on the first
and third sheets, $\infty_{1,3}$ where it behaves locally as

\begin{align}
h_3(Q)&=\begin{cases}
\mathrm{ln}\,z - \mathrm{ln}\delta_3+o(1) &\text{at}\quad
Q\longrightarrow \infty_3,\\
-\mathrm{ln}\,z + \mathrm{ln}\delta_3+o(1) & \text{at}\quad
Q\longrightarrow \infty_1.
\end{cases}\label{h3}
\end{align}
In (\ref{h2}) and (\ref{h3}) $\delta_{2}$ and $\delta_3$ are certain
constants.
\end{definition}

Observe that the $\mathfrak b$-periods of integrals  $h_2(Q), h_3(Q)$
are given as
\[ \frac{1}{2\pi \imath}\oint_{{\mathfrak b}_j}
dh_2 = \int\limits_{\infty_2}^{\infty_1}dv_j,\qquad
\frac{1}{2\pi \imath}\oint_{{\mathfrak b}_j} dh_3
=\int\limits_{\infty_3}^{\infty_1}dv_j,\quad
j=1,\ldots,g.
 \]
Denote these periods as $- \boldsymbol{r}_{2,3}$, so that
\begin{equation} \int\limits_{\infty_1}^{\infty_{2,3}}d\boldsymbol{v}
=\boldsymbol{r}_{2,3}. \label{rij}
\end{equation}

Taking into account that $\sigma(\infty_{j}) = \infty_{j}$ (cf.
(\ref{sheets}))
and, once again, (\ref{vcon}), we conclude that
\begin{equation}\label{rcon}
\overline{\boldsymbol{r}_{2,3}} = - \boldsymbol{r}_{2,3}
\quad \mbox{mod}\,\, {\Bbb Z}.
\end{equation}

The $\sigma$ - invariance of the leading terms of the asymptotic of
the differentials $dh_{2}$, $dh_{3}$ at the points $\infty_{k}$ (cf.
(\ref{h2}) and  (\ref{h3})) implies, instead of (\ref{vcon}), the
symmetry equations,
\begin{equation}\label{hsym}
\sigma^{*}dh_{2,3} = \overline{dh_{2,3}}.
\end{equation}
Similar to (\ref{E1}) - (\ref{N2}), the constant terms in the
asymptotic (\ref{h2}) - (\ref{h3}) can be determined with the help
of the relations
\begin{equation}\label{deltak}
\int_{Q_{k}}^{Q_{1}}dh_{k} \equiv \oint_{\mathfrak{l}_{k}}dh_{k}
-2\ln z + 2\ln \delta_{k} + O\left(\frac{1}{z}\right),
\end{equation}
$$
z \to + \infty,
\quad k = 2, 3.
$$
Similar to the case of the integrals $\Omega_{1, 2}$, we can
now use the symmetry properties (\ref{involl2}), (\ref{involl3}),
and (\ref{hsym}) to see  that
\begin{equation}\label{deltasym}
\overline{\delta_{2,3}} = -\delta_{2,3}\,.
\end{equation}
Indeed, for $k =2, 3$ we have
$$
\overline{\oint_{\mathfrak{l}_{k}}dh_{k}}=
\oint_{\mathfrak{l}_{k}}\overline{dh_{k}} =
\oint_{\mathfrak{l}_{k}}\sigma^{*}dh_{k}
$$
$$
= \oint_{\sigma(\mathfrak{l}_{k})}dh_{k} =
\oint_{\mathfrak{l}_{k}}dh_{k} +2\pi \imath
m^{(k)}\mbox{res}|_{\infty_{1}}dh_{k} +2\pi \imath
n^{(k)}\mbox{res}|_{\infty_{k}}dh_{k}
$$
$$
=\oint_{\mathfrak{l}_{k}}dh_{k} + 2\pi \imath (m^{(k)} - n^{(k)}),
$$
and (\ref{deltasym}) follows in virtue of the asymptotic relation
(\ref{deltak}) and the parity relation (\ref{odddif} ) ( equation
(\ref{odddif}) is now important - it is responsible for the minus
sign in (\ref{deltasym})).

In the next section we shall construct explicitly the
integrals $\Omega_{1,2}(P)$ and $h_{2,3}(P)$ and compute
the constants $E_{1,2},N_{1,2},\delta_{2,3}$ in terms of
$\theta$-functions of the curve $X$.

\subsection{$\theta$-function and prime-form}
The $\theta$-function of the curve $X$
with characteristic $[\varepsilon]$
\[
 [\varepsilon] = \left[\begin{array}{c}
{\boldsymbol{\varepsilon}'}^T\\
\boldsymbol{\varepsilon}^T
\end{array}\right]
=\left[\begin{array}{ccc}
\varepsilon_1'&\ldots &\varepsilon_g'\\
\varepsilon_1 &\ldots &\varepsilon_g
\end{array}
\right]
\]
is given by the formula
\begin{align}
\theta[\varepsilon](\boldsymbol{v})&= \sum_{{\boldsymbol n}\in{\mathbb
Z}^g} \mathrm{exp} \left\{ \imath\pi
(\boldsymbol{n}+\boldsymbol{\varepsilon}')^T\tau(\boldsymbol{n}
+\boldsymbol{\varepsilon}')+2\imath\pi(\boldsymbol{n}+
\boldsymbol{\varepsilon}')^T(\boldsymbol{v}
+\boldsymbol{\varepsilon}) \right\}. \label{theta}
\end{align}
In this paper we are considering only half-integer characteristics,
$\varepsilon_k', \varepsilon_l= 0$ or $\frac12$ for any
$k,l=1,\ldots,g$. Even characteristic $[\varepsilon]$
($e^{4i\pi\boldsymbol{\varepsilon}^T
\boldsymbol{\varepsilon}'} = 1$) is nonsingular
if $\theta[\varepsilon](\boldsymbol{0})\neq 0$. Odd characteristic
($e^{4i\pi\boldsymbol{\varepsilon}^T
\boldsymbol{\varepsilon}'} = - 1$) is
nonsingular if the gradient
$ \nabla \theta[\varepsilon](\boldsymbol{0}) =(
\frac{\partial}{\partial v_1}
\theta[\varepsilon](\boldsymbol{v})\big|_{\boldsymbol{v}=0},\ldots,
\frac{\partial}{\partial v_g}
\theta[\varepsilon](\boldsymbol{v})\big|_{\boldsymbol{v}=0}
)^T $ is non-zero.

The canonical $\theta$-function is the $\theta$-function with zero
characteristic
\begin{equation}\label{thetacan}
\theta(\boldsymbol{v})=\sum_{{\boldsymbol n}\in{\mathbb Z}^g}
\mathrm{exp}\left\{ \imath\pi
 \boldsymbol{n}^T\tau\boldsymbol{n}
 +2\imath\pi\boldsymbol{n}^T\boldsymbol{v}  \right\}.
\end{equation}

The $\theta$-function with a characteristic $[\varepsilon]$ possesses
the periodicity property:
\begin{align}\begin{split}
\theta[\varepsilon](\boldsymbol{v}+\boldsymbol{e}_k)&
=\mathrm{exp}\{ -2\imath \pi
\varepsilon'_k\}\theta[\varepsilon](\boldsymbol{v})
,\\
\theta[\varepsilon](\boldsymbol{v}+\boldsymbol{\tau}_k)
&=\mathrm{exp}\{-\imath\pi\tau_{kk}-2\imath\pi
v_k  -2\imath \pi \varepsilon_k  \}\theta[\varepsilon](\boldsymbol{v}),
\end{split}\label{periodicity}
\end{align}
where $\boldsymbol{e}_k=(0,\ldots,1,\ldots,0)^T$,
$\boldsymbol{\tau}_k=(\tau_{1k},\ldots,
\tau_{gk})^T$ and $k=1,\ldots,g$.
Using the relationship between $\overline{\tau}$ and $\tau$ derived
above (see (\ref{tauinv})) and definition of the $\theta$-function,
we have that
\begin{equation}\label{thetasym1}
\overline{\theta[\varepsilon](\boldsymbol{v})}e^{-\imath \pi{\boldsymbol{\varepsilon}'}^T\tau_{0}\boldsymbol{\varepsilon}'}
\theta[\varepsilon](\overline{\boldsymbol{v}}),
\end{equation}
and for the canonical $\theta$ - function,
\begin{equation}\label{thetasym2}
\overline{\theta(\boldsymbol{v})}=\theta(\overline{\boldsymbol{v}}).
\end{equation}

The Schottky-Klein prime form \cite{ba97,fa73} is defined everywhere
on $X\times X$ and is introduced
by the formula
\begin{equation}
{\mathcal E}(P,Q)=\frac{\theta[\varepsilon]\left(\int\limits_{Q}^{P}
d\boldsymbol{v} \right)}
{
\sqrt{\sum_{k=1}^g
\frac{\partial \theta[\varepsilon](\boldsymbol{0})}
{\partial v_k}dv_k(P) }
\sqrt{\sum_{k=1}^g
\frac{\partial \theta[\varepsilon](\boldsymbol{0})}
{\partial v_k}dv_k(Q) }
},
\label{skpf}
\end{equation}
where $P=(x,y)\in X$ and $Q=(z,w)\in X$ are arbitrary points and
$\theta[\varepsilon](\boldsymbol{v})$ is the $\theta$-function,
with non-singular odd half-integer characteristic $[\varepsilon]$.
Concerning the
characteristic $[\varepsilon]$ it is natural to suppose without
loosing generality that the
vector
\[ \boldsymbol{e}=\boldsymbol{\varepsilon}+\tau
\boldsymbol{\varepsilon}'\in (\theta)\subset \mathrm{Jac} (X), \quad
\theta(\boldsymbol{e})=0,
\]
where $(\theta)$ be $\theta$-divisor, is parametrized as
\begin{equation} \boldsymbol{e}=\sum_{k=1}^{g-1}\int\limits_{P}^{P_k}
d\boldsymbol{v}-\boldsymbol{K}_{P},\end{equation}
where $\boldsymbol{K}_{P}$ is vector of Riemann constants with base
point $P$ and points $P_1,\ldots,P_{g-1}$ are different branch points of
the curve $X$.

The prime-form
${\mathcal E}(P,Q)$ vanishes only on the diagonal, $P=Q$, in
the vicinity of which it is expanded in power series as
\begin{equation}{\mathcal E}(P,Q)\sqrt{d \xi(P) d \xi(Q)}
=\xi(P)-\xi(Q)  +  O(\xi(P)-\xi(Q))
\label{eexpan},\end{equation}
where $\xi(P)$ and $\xi(Q)$ are local coordinates of the points $P$ and $Q$
around $P_0$, $\xi(P_0)=0$.

The prime-form (\ref{skpf}) permits to construct symmetric
second kind differential 2-differential which is called  Bergmann kernel
on $X\times X$ as
\begin{align} d\omega(P,Q)=&\frac{\partial^2}{\partial x\partial z}
\;\mathrm{ln}\,{\mathcal E}(P,Q)\,dxdz\label{bkernel}\\
&=\frac{\partial^2}{\partial x\partial z}
\;\mathrm{ln}\,\theta[\varepsilon]\left(\int\limits_{Q}^{P}
d\boldsymbol{v} \right)\,dxdz,\notag
\end{align}
where $[\varepsilon]$ are non-singular odd half-integer characteristics.

The differential $d\omega(P,Q)$, where the coordinates $P,Q$ are
given as $P=(x,y)$, $Q=(z,w)$ has the properties:

{\bf i)} It is symmetric,  $d\omega(P,Q)=d\omega(Q,P)$.

{\bf ii)} It is holomorphic except on the diagonal set ($P=Q$) where
it has a double pole. If the points $P,Q$ are places in the vicinity
of the point $P_0$ and  $\xi$ is the local coordinate
around $P_0$, $\xi(P_0)=0$ then the expansion of
$d\omega(P,Q)$ near $P_0$ takes the form

\begin{align} d\omega(P,Q)
=\left(\frac{1}{(\xi(P)-\xi(Q))^2}+
\frac16 S(P_0)+ O(\xi(P)-\xi(Q))\right)
d\xi(P)d\xi(Q),\label{bkexp}
\end{align}
where $S(P_0)$ holomorphic projective connection (explicit expression in
terms of $\theta$-functions is given e.g. in  \cite{fa73}).

{\bf iii)} The $\mathfrak {a}$-periods taken in variable $P$ or $Q$
vanish,
\begin{equation}\oint\limits_{\mathfrak{a}_i}d\omega(P,Q)=0,
\quad i=1,\ldots,g.
\end{equation}

Introduce following notations for the directional derivatives:
\begin{align*}
\partial_{\boldsymbol{V}}f(\boldsymbol{v})=&\sum_{k=1}^gV_k\frac{\partial}{\partial
v_k}f(\boldsymbol{v}),\\
\partial^2_{\boldsymbol{V},\boldsymbol{W}}f(\boldsymbol{v})&\sum_{k=1}^g\sum_{l=1}^gV_kW_l
\frac{\partial^2}{\partial v_k\partial v_l}f(\boldsymbol{v}),\;\text{etc.},
\end{align*}
where $\boldsymbol{V}=(V_1,\ldots,V_g)^T$,
$\boldsymbol{W}=(W_1,\ldots,W_g)^T$ are constant vectors and
$f(\boldsymbol{v})$ is a function of the vector argument
$\boldsymbol{v}=(v_1,\ldots,v_g)^T$. The following theorem can be
found in \cite{fa73},\cite{jo92} concerning directional derivatives
along the $\theta$-divisor.

\begin{theorem} For any nonzero vectors $\boldsymbol{a}=(a_1,\ldots,a_g)^T$,
 $\boldsymbol{b}=(b_1,\ldots,b_g)^T \in \mathbb C^g$ and points
$P_1,\ldots,P_{g-1}$ on $X$ the following identity holds

\begin{equation}
\frac{\sum\limits_{j=1}^g \frac{\partial \theta}{\partial v_j}
(\boldsymbol{e})a_j}{\sum\limits_{j=1}^g
 \frac{\partial \theta}{\partial v_j}(\boldsymbol{e})b_j}
=\frac{\mathrm{det}\left[ \boldsymbol{a}|
{d}\boldsymbol{v}(P_1)|\ldots |
{d}\boldsymbol{v}(P_{g-1})\right]  }{\mathrm{det}\left[ \boldsymbol{b}|
{d}\boldsymbol{v}(P_1)|\ldots |
{d}\boldsymbol{v}(P_{g-1}) \right]  },\label{jorgenson}
\end{equation}
where the point $\boldsymbol{e}$ is given by
\begin{equation} \boldsymbol{e} = \sum_{k=1}^{g-1}
\int\limits_{P}^{P_k}\mathrm{d\boldsymbol{v}} -\boldsymbol{K}_P
\label{pointe} \end{equation}
and the matrices in (\ref{jorgenson}) have been expressed by
indicating each of $g$ columns.
\end{theorem}

\begin{lemma} Let
  $[\varepsilon]$ be non-singular odd half-integer characteristic
of the curve $X$. Then
\begin{align} &
  \theta[\varepsilon]\left(\int\limits_{\infty_i}^{\infty_j}
d\boldsymbol{v}\right)
\neq 0,\quad i\neq j=1,2,3\label{tr23},\\
& \partial_{\boldsymbol{V}^{(i)}}\theta[\varepsilon](\boldsymbol{0})\neq
 0,\quad i=1,2,3.\label{dtr23}
    \end{align}
\end{lemma}
\begin{proof}
The holomorphic differential $\mathrm{d}u_1=\imath \mathrm{d}z/f_w$
vanishes at $\infty_i$, $i=1,2,3$ to the order $4n-8=2g-2$ and
therefore  $(2g-2)\infty_i$ is equivalent to the canonical class.
Hence the vector of Riemann constants with the base point at
$\infty_i$ is a half-period \cite{fk80}. The curve considered has
only simple branch points (see the proof of Lemma \ref{genuslemma})
and therefore non-singular odd half-period corresponding to the
characteristic $[\varepsilon]$ can be given by the formula
(\ref{pointe}) with $P_i$, $i=1,\ldots$ being branch points.

First prove (\ref{dtr23}) for $i=1$, $j=2$.
Suppose the opposite. Then the vanishing of the
directional derivative
$\partial_{\boldsymbol{V}^{(2)}}\theta(\boldsymbol{0})=0$ will lead,
according to (\ref{jorgenson}), to the vanishing of the determinant
\[\mathrm{det}\left[ \boldsymbol{V}^{(2)}|
{d}\boldsymbol{v}(P_1)|\ldots |
{d}\boldsymbol{v}(P_{g-1})\right]=\mathrm{det}
\left[\left.\left.\left.\frac{{d}\boldsymbol{v}(\infty_2)}{d
\xi(\infty_2)}   \right|
{d}\boldsymbol{v}(P_1)\right|\ldots \right|
{d}\boldsymbol{v}(P_{g-1})\right]
\]
But there is no $\infty_2$ among branch points $P_i$ as that was shown
earlier. The contradiction obtained proves the statement. Other cases
are considered analogously.

Prove (\ref{tr23}). Suppose the opposite.
Consider further the prime-form $\mathcal{E}(P,\infty_1)$ given by the
formula
(\ref{skpf}). It is well defined because directional derivative
$\partial_{\boldsymbol{V}^{(1)}}\theta(\boldsymbol{0})\neq 0$.
According to principal
property of the prime-form it vanishes only at $P=\infty_1$. Therefore
the supposed vanishing of the $\theta$-function should lead to
vanishing of the directional derivative in the denominator. The
contradiction obtained proves the statement.
\end{proof}

\subsection{$\theta$-functional construction of meromorphic integrals}
To construct the required second and third kind integrals
$\Omega_{1,2}(Q)$ and $h_{2,3}(Q)$ we first construct
corresponding meromorphic differentials with the aid of
prime-form introduced.

The normalized meromophic differential of the third kind,
$dh_2$ with
the poles in $\infty_1$ and $\infty_2$ of the first order and residues
$\pm 1$ in the poles  is given as
\begin{equation}
dh_2(P) = d_z \,
\mathrm{ln}\,\frac{{\mathcal E}(P,\infty_1)}{{\mathcal E}(P,\infty_2)}.
\end{equation}
Analogously the normalized meromophic differential of the third kind,
$dh_3$ with
the poles in $\infty_1$ and $\infty_3$ the first order and residues
$\pm 1$ in the poles  is given as
\begin{equation}
dh_3(P) = d_z \,
\mathrm{ln}\,\frac{{\mathcal E}(P,\infty_1)}{{\mathcal E}(P,\infty_3)}.
\end{equation}

We are in position now to give $\theta$-functional representation
for the second and third kind integrals, which permit us to compute
6 constants $E_{1,2}$, $N_{1,2}$ and $\delta_{2,3}$ in terms of
$\theta$-functions.

Consider three quantities
\begin{align}\begin{split}
\Omega^{(i)}_1(P)&=\left.\int  dx\left\{
\frac{\partial}{\partial x\partial z}
      \mathrm{ln}\,\theta[\varepsilon]\left(\int\limits_{Q}^P
      d\boldsymbol{v}\right) dz \right\} \right|_{Q=\infty_i}
      \frac{1}{d \xi(\infty_i)}\\
&=\left.\sum_{k=1}^gV_k^{(i)}\frac{\partial}{\partial v_k}\,\mathrm{ln}\,
\theta[\varepsilon](\boldsymbol{v})\right|_{\boldsymbol{v}=\int\limits_{\infty_i}^P
d\boldsymbol{v}}\\
&=\partial_{\boldsymbol{V}^{(i)}}\,\mathrm{ln}\,
\theta[\varepsilon]\left(\int\limits_{\infty_i}^P d \boldsymbol{v}  \right),
\end{split}\label{omega11}\end{align}
where $i=1,2,3$ and the point $Q$ has coordinates  $(z,w)$.

\begin{lemma} The quantities $\Omega^{(i)}_1(P)$, $i=1,2,3$ are second
  kind Abelian integrals with unique pole of the first order
  at corresponding to the
  index $i$ infinity $\infty_i$ with $\boldsymbol{\mathfrak{a}}$ and
$\boldsymbol{\mathfrak{b}}$-periods,

\begin{equation}
\oint\limits_{\mathfrak{a}_l} d \Omega^{(i)}_1(P) =0,\quad
\oint\limits_{\mathfrak{b}_l} d \Omega^{(i)}_1(P)  =-2\imath\pi
V_l^{(i)}\quad l=1,\ldots,g\label{periods}
 \end{equation}
and following behaviour at the
infinities $\infty_1,\infty_2,\infty_3$ on different sheets
\begin{align}\label{expomega1}\begin{split}
\Omega_1^{(1)}(P)&=\begin{cases}\frac{1}{\xi} +        c_1^{(1)}+O(\xi),\\
                             X_{1,2}+O(\xi),\\
                             X_{1,3}+O(\xi), \end{cases}\qquad
\Omega_1^{(2)}(P)=\begin{cases}X_{2,1}+O(\xi),\\
                            \frac{1}{\xi} +c_1^{(2)}+O(\xi),\\
                             X_{2,3}+O(\xi), \end{cases}\\
\Omega_1^{(3)}(P)&=\begin{cases}X_{3,1}+O(\xi),\\
                             X_{3,2}+O(\xi),\\
                             \frac{1}{\xi} + c_1^{(3)}+O(\xi),
\end{cases}\end{split}
\end{align}
where
\begin{align}
X_{i,j}&=\partial_{\boldsymbol{V}^{(i)}}\mathrm{ln}\theta[\varepsilon]
\left(\int\limits_{\infty_i}^{\infty_j}d
  \boldsymbol{v}\right),\quad i\neq j=1,2,3, \label{xij}\\
c_1^{(i)}&=\frac{1}{2}\frac{\partial^{2}_{\boldsymbol{V}^{(i)},\boldsymbol{V}^{(i)}}
\theta[\varepsilon](\boldsymbol{0})
} {\partial_{\boldsymbol{V}^{(i)}}
  \theta[\varepsilon](\boldsymbol{0})}
-\frac{\partial_{\boldsymbol{W}^{(i)}} \theta[\varepsilon](\boldsymbol{0}) }
{\partial_{\boldsymbol{V}^{(i)}}
  \theta[\varepsilon](\boldsymbol{0})},\quad i=1,2,3.\label{c1i}
\end{align}
\end{lemma}
\begin{proof}
Relations (\ref{periods}) for periods follow immediately from
definition and periodicity properties (\ref{periodicity})
of the $\theta$-function, while expansions (\ref{expomega1}) result
direct computation. The quantities (\ref{xij}) and (\ref{c1i}) are
well defined because of inequalities (\ref{tr23}) and (\ref{dtr23}).
\end{proof}

Similar statement is valid for three quantities
\begin{align}
\Omega^{(i)}_2(P) = \left(2\partial_{\boldsymbol{W}^{(i)}} -
\partial^2_{\boldsymbol{V}^{(i)}, \boldsymbol{V}^{(i)}  }\right)\mathrm{ln}\,
\theta[\varepsilon]\left(\int\limits_{\infty_i}^P
d \boldsymbol{v}  \right)   ,\label{omega22}\end{align}
where $i=1,2,3$ and the point $Q$ has coordinates  $(z,w)$.

\begin{lemma} The quantities $\Omega^{(i)}_2(P)$, $i=1,2,3$ are second
  kind Abelian integrals with unique pole of second order
  at corresponding to the
  index $i$ infinity $\infty_i$ with $\boldsymbol{\mathfrak{a}}$ and
$\boldsymbol{\mathfrak{b}}$-periods,

\begin{equation}
\int\limits_{\mathfrak{a}_l} d \Omega^{(i)}_2(P) =0,\quad
\int\limits_{\mathfrak{b}_l} d \Omega^{(i)}_2(P)  =-4\imath\pi
W_l^{(i)}\quad l=1,\ldots,g\label{periods2}
 \end{equation}
and following behavior at the infinities on different sheets
\begin{align}\label{expomega2}\begin{split}
\Omega_2^{(1)}(P)&=\begin{cases}\frac{1}{\xi^2} +        c_2^{(1)}+O(\xi),\\
                             Y_{1,2}+O(\xi),\\
                             Y_{1,3}+O(\xi), \end{cases}\qquad
\Omega_2^{(2)}(P)=\begin{cases}Y_{2,1}+O(\xi),\\
                            \frac{1}{\xi^2} +c_2^{(2)}+O(\xi),\\
                             Y_{2,3}+O(\xi), \end{cases}\\
\Omega_2^{(3)}(P)&=\begin{cases}Y_{3,1}+O(\xi),\\
                             Y_{3,2}+O(\xi),\\
                             \frac{1}{\xi^2} + c_2^{(3)}+O(\xi),
\end{cases}\end{split}
\end{align}
where
\begin{align}
Y_{i,j}&=\left(2\partial_{\boldsymbol{W}^{(i)}} -
\partial^2_{\boldsymbol{V}^{(i)}, \boldsymbol{V}^{(i)}  }\right)
\mathrm{ln}\theta[\varepsilon]\left(\int\limits_{\infty_i}^{\infty_j}d
  \boldsymbol{v}\right),\label{yij}\\
c_2^{(i)}&=-\left(\frac{\partial_{\boldsymbol{W}^{(i)}}
\theta[\varepsilon](\boldsymbol{0}) }
{\partial_{\boldsymbol{V}^{(i)}}
\theta[\varepsilon](\boldsymbol{0})}\right)^2
+\frac{1}{3}\frac{\partial^3_{\boldsymbol{V}^{(i)},
\boldsymbol{V}^{(i)},\boldsymbol{V}^{(i)}}
 \theta[\varepsilon](\boldsymbol{0}) }
{\partial_{\boldsymbol{V}^{(i)}} \theta[\varepsilon](\boldsymbol{0})}
+\frac{\partial_{\boldsymbol{Z}^{(i)}} \theta[\varepsilon](\boldsymbol{0}) }
{\partial_{\boldsymbol{V}^{(i)}} \theta[\varepsilon](\boldsymbol{0})}
\label{c2i}.
\end{align}

\end{lemma}
\begin{proof} The second term in (\ref{omega22}) does not contribute to
  periods -- it is meromorphic function but the first term  in
(\ref{omega22})
leads to the relations (\ref{periods2}).
The expansions (\ref{expomega2}) result direct computation.  The
quantities (\ref{yij}) and (\ref{c2i}) are
well defined because of inequalities (\ref{tr23}) and (\ref{dtr23}).
\end{proof}

The normalized meromorphic differential of
the second kind $d\Omega_1(P)$ and  $d\Omega_2(P)$  are then given as
\begin{align}
\Omega_1(P)&=-\imath\Omega_1^{(1)}(P)+\imath\Omega_1^{(2)}(P)+\imath\Omega_1^{(3)}(P)+C_1
             \label{domega1},\\
\Omega_2(P)&=-2\imath\Omega_2^{(1)}(P)+2\imath\Omega_2^{(2)}(P)+2\imath\Omega_2^{(3)}(P)+C_2,
             \label{domega2}
\end{align}
where $C_k$, $k=1,2$ are constants and integrals $\Omega_i^{(j)}(P)$
are given in (\ref{omega11}) and (\ref{omega22}).

The representations (\ref{domega1},\ref{domega2}) of
the meromorphic differentials permits to compute
the constants $E_{1,2}, N_{1,2}$ and $\delta_{2,3}$
in terms of $\theta$-functions

\begin{theorem}\label{6constants}
The following $\theta$-functional expressions are valid for the
constants $E_{1,2}, N_{1,2}$ and $\delta_{2,3}$:
\begin{align}
\begin{split}
E_1&=\imath\left(X_{12}+X_{21}+X_{31}-X_{32}-c_1^{(2)}-c_1^{(1)}\right),\\
E_2&=\imath\left(X_{13}+X_{31}+X_{21}-X_{23}-c_1^{(3)}-c_1^{(1)}\right)
\end{split}\label{e12}
\end{align}
and
\begin{align}\begin{split}
N_1&=-2\imath\left(Y_{12}+Y_{21}+Y_{31}-Y_{32}-c_2^{(2)}-c_2^{(1)}\right),\\
N_2&=-2\imath\left(Y_{13}+Y_{31}+Y_{21}-Y_{23}-c_2^{(3)}-c_2^{(1)}
\right)\end{split}\label{n12}
\end{align}
and
\begin{align}
\delta_{2,3}&=\frac{\imath}
{\theta[\varepsilon](\boldsymbol{r}_{2,3})}\sqrt{
\partial_{\boldsymbol{V}^{(2,3)}}
\theta[\varepsilon](\boldsymbol{0})}\sqrt{
\partial_{\boldsymbol{V}^{(1)}}
\theta[\varepsilon](\boldsymbol{0})}
   \label{delta23}
,\end{align}
where quantities $X_{i,j},Y_{i,j},c_{k}^{(i)} $ are defined in
(\ref{xij}), (\ref{yij}), (\ref{c1i}) and (\ref{c2i}).
\end{theorem}
\begin{proof}
Consider first integral with the first order poles at infinities.
Expand (\ref{domega1}) at $\infty_1,\infty_2,\infty_3$
and compare with the asymptotic conditions (\ref{omega1}) to obtain equations
\begin{align*}
\imath X_{21}+\imath X_{31}+C_1-\imath c_1^{(1)}
&=\frac12E_1+\frac12E_2,\\
-\imath X_{12}+\imath X_{32}+C_1+\imath c_1^{(2)}&=\frac12E_2-\frac12E_1,\\
-\imath X_{13}+\imath X_{23}+C_1+\imath c_1^{(3)}&
=-\frac12E_2+\frac12E_1.
\end{align*}
We find (\ref{e12}) and the following expression for the constant
\begin{align*}
C_1&=\frac{\imath}{2}\left(X_{12}+X_{13}\right)-\frac{\imath}{2}\left(X_{32}+X_{23}\right)+
\frac{\imath}{2}(c_1^{(2)}+c_1^{(3)}).
\end{align*}

Consider further the integral with the second order poles at infinities.
Expand (\ref{domega2}) at $\infty_1,\infty_2,\infty_3$
and compare with the asymptotic conditions (\ref{omega2}). Solving
linear equations as before we find (\ref{n12}) and the following
expression for the constant
\begin{align*}
C_2&=\frac{\imath}{2}\left(Y_{12}+Y_{13}\right)-\frac{\imath}{2}\left(Y_{32}+Y_{23}\right)+
\frac{\imath}{2}(c_2^{(2)}+c_2^{(3)}).
\end{align*}

Consider further the third kind integral
\begin{equation}\label{int3k}
h_2(P)=\mathrm{ln}\,\frac{{\mathcal E}(P,\infty_1)}
{{\mathcal E}(P,\infty_2)}+C_h.
\end{equation}
On the first sheet we have that, as $P \to \infty_{1}$,

\begin{align}
h_2(P)
&=\mathrm{ln}\frac{\theta[\varepsilon]\left(\int\limits_{\infty_1}^P d
    \boldsymbol{v}\right)}{\theta[\varepsilon]\left(\int\limits_{\infty_2}^P
d
 \boldsymbol{v}\right)}\frac{ \sqrt{\sum_{k=1}^g
\frac{\partial\theta[\varepsilon](\boldsymbol{0})}{\partial
    v_k} d v_k(\infty_2)}}{\sqrt{\sum_{k=1}^g
\frac{\partial\theta[\varepsilon](\boldsymbol{0})}{\partial
    v_k} d v_k(\infty_1)}} +C_h +O(\xi)  \notag \\
&=\mathrm{ln}\,\xi+\mathrm{ln}
\sqrt{\sum_{k=1}^gV_k^{(2)}\frac{\partial}{\partial v_k}\,
\theta[\varepsilon](\boldsymbol{0})}
\sqrt{\sum_{k=1}^gV_k^{(1)}
\frac{\partial}{\partial v_k}\,
\theta[\varepsilon](\boldsymbol{0})}
\label{exph22}
\\& - \mathrm{ln}\,
\theta[\varepsilon](-\boldsymbol{r}_2)+C_h+O(\xi)\notag
\end{align}
whilst on the second sheet, as $P \to \infty_{2}$, we have
\begin{align}
h_2(P)
&=-\mathrm{ln}\,\xi-\mathrm{ln}
\sqrt{\sum_{k=1}^gV_k^{(2)}\frac{\partial}{\partial v_k}\,
\theta[\varepsilon](\boldsymbol{0})}
\sqrt{\sum_{k=1}^gV_k^{(1)}
\frac{\partial}{\partial v_k}\,
\theta[\varepsilon](\boldsymbol{0})}
\notag \\
&+ \mathrm{ln}\,\theta[\varepsilon](\boldsymbol{r}_2)+C_h+O(\xi)
\label{exph21}.
\end{align}

Comparison  of expansions (\ref{exph22}) and
(\ref{exph21}) with the expansions (\ref{h2})  leads to
$C_h=\frac{\imath\pi}{2}$, and
\[ \delta_2=\frac{\imath}{\theta[\varepsilon](\boldsymbol{r}_{2})}
\sqrt{\sum_{k=1}^gV_k^{(2)}\frac{\partial}{\partial v_k}\,
\theta[\varepsilon](\boldsymbol{0})}
\sqrt{\sum_{k=1}^gV_k^{(1)}
\frac{\partial}{\partial v_k}\,
\theta[\varepsilon](\boldsymbol{0})}.\]

The expression for $\delta_2$ in (\ref{delta23}) follows.
The expression for $\delta_3$ in (\ref{delta23}) is derived analogously.
\end{proof}

We emphasize that the constants described in the Theorem
\ref{6constants}
are fundamental: expression for the constants $c_2^{(i)}$
coincide with accuracy to a trivial multiplier with
values of the projective connection, $S(P)$ (see (\ref{bkexp})) at
infinities,  $S(\infty_i), i=1,2,3$. The
quantities $X_{i,j}$ and $Y_{i,j}$ can be expressed in terms of
multidimensional Kleinian $\zeta$ and $\wp$-function whose classical
and modern treatment, in the hyperelliptic case,
can be found in \cite{ba97} and \cite{bel97b} correspondingly.
We also remark that analogous expressions for constants $E_{1,2}$, $N_{1,2}$
in terms of $\theta$-functions and winding vectors for Thirring model
which is associated with a hyperelliptic curve are obtained in
\cite{egh00}, see also \cite{gh03}.

The important symmetry relations for the costants  $E_{1,2}$, $N_{1,2}$,
and $\delta_{2,3}$, which have been  derived earlier
(see (\ref{ENcon}) and (\ref{deltasym})), can be also easily
obtained from the $\theta$-functional formulae of
Theorem \ref{6constants} with the help of equation (\ref{thetasym1}).

\section{Algebro-geometric solutions of the Manakov system}

We now summarize a list of basic objects which are related to the
curve (\ref{trigonal})

{\bf 1.} A homology basis of oriented cycles $\mathfrak{a}_j$ and
$\mathfrak{b}_j$
as discussed in the Section 3.

{\bf 2.} The differentials $dv_j$ introduced in the Section 4.1
are normalized.

{\bf 3.} The matrix of the $\mathfrak {b}$-periods of the trigonal
curve $X$ and the associated $\theta$-functions as defined by
equations (\ref{tau}) and (\ref{theta}) respectively.

{\bf 4.} The Abelian integrals $\Omega_1(Q)$, $\Omega_2(Q)$,
$h_1(Q)$ and $h_2(Q)$, $Q\in X$ which are fixed by the conditions
(\ref{omega1}), (\ref{omega2}) and (\ref{h2}), (\ref{h3}).

{\bf 5.} An arbitrary divisor $\mathcal D$ with degree $\mathrm{deg}
{\mathcal
D} =g$ of general position, i.e.
$$\mathcal{ D}= \sum_{i=1}^g Q_i, \quad \pi(Q_{i})\neq e_i, \quad i\neq k
\Rightarrow
\pi(Q_{i})\neq \pi(Q_j),
$$
where $\pi(P)$ is three sheeted covering
\begin{equation*}
\pi: X\longrightarrow \mathbb{CP}^1,\qquad
\pi^{-1}(\infty)=(\infty_1,\infty_2,\infty_3),
\end{equation*}
and $e_i$ are the branch points of the curve $X$.

The vector valued Baker-Akhiezer function
$$\boldsymbol{\Psi}(Q,x,t)= (\psi_1(Q,x,t),  \psi_2(Q,x,t),
 \psi_3(Q,x,t))^T$$
is uniquely defined by two conditions. The first of these
conditions describes the analytic structure of $\boldsymbol{\Psi}$
on $X/\{\infty_{1,2,3}\}$

{\bf I.} $\psi_i(Q,x,t)$, $i=1,2,3$ are meromorphic on
$X/\{\infty_{1,2,3}\}$. Their
divisor of poles coincides with $\mathcal D$.

The second condition describes the asymptotic behavior of
$\boldsymbol{\Psi}(Q,x,t)$ at $\infty_{1,2,3}$ and shows that
$\boldsymbol{\Psi}(Q,x,t)$ has essential singularities at
$\infty_{1,2,3}$

{\bf II.} As $Q \to \infty_{1,2,3}$, the
asymptotic behavior of $\boldsymbol{\Psi}(Q,x,t)$
is given by the equations,
\begin{eqnarray*}
\boldsymbol{\Psi}(Q)&=&\left[\left(\begin{array}{c}1\\0\\0
\end{array}\right)
+O(z^{-1})\right]\mathrm{exp}(-\imath z x -2\imath z^2 t)\\
                    &\text{at}&  Q\longrightarrow \infty_1,\quad z=\pi(Q),\\
\boldsymbol{\Psi}(Q)&=&\frac{z}{\delta_2}\left[\left(\begin{array}{c}0\\1\\0
\end{array}\right)
+O(z^{-1})\right]\mathrm{exp}(\imath z x +2\imath z^2 t)\\
                    &\text{at}&  Q\longrightarrow \infty_2,\quad z=\pi(Q),\\
\boldsymbol{\Psi}(Q)&=&\frac{z}{\delta_3}\left[\left(\begin{array}{c}0\\0\\1
\end{array}
\right)
+O(z^{-1})\right]\mathrm{exp}(\imath z x +2\imath z^2 t)\\
                    &\text{at}&  Q\longrightarrow \infty_3,\quad z=\pi(Q)
,\end{eqnarray*}
where $\delta_2,\delta_3\in\mathbb C$ are non-zero constants. Indeed,
we shall take $\delta_{2}$ and $\delta_{3}$ from (\ref{h2}) and
(\ref{h3}), respectively.

Then,
$\boldsymbol{\Psi}(Q,x,t)=(\psi_1(Q,x,t),\psi_2(Q,x,t),\psi_3(Q,x,t))^T$
is uniquely determined by the conditions {\bf I.} and {\bf II.}
and may be explicitly constructed by the formula
\begin{align}
\psi_1(Q)&=\frac{\theta\left(\int\limits_{\infty_1}^Qd\boldsymbol{v}+\boldsymbol{\Gamma}
\right)
                       \theta(\boldsymbol{D}) }
              {\theta\left(\int\limits_{\infty_1}^Qd\boldsymbol{v}-\boldsymbol{D}
\right)
              \theta(\boldsymbol{\Gamma}) }\mathrm{exp}\{x\Omega_1(Q)
              +t\Omega_2(Q)-Ex+Nt  \},
              \label{BAf1}\\
\psi_2(Q)&=\frac{\theta\left(\int\limits_{\infty_1}^Qd\boldsymbol{v}+\boldsymbol{\Gamma}
-\boldsymbol{r}_2 \right)
                       \theta(\boldsymbol{D}-\boldsymbol{r}_2) }
              {\theta\left(\int\limits_{\infty_1}^Qd\boldsymbol{v}-\boldsymbol{D}
\right)
              \theta(\boldsymbol{\Gamma}) }
              \notag\\&\qquad\times
              \mathrm{exp}\{x\Omega_1(Q)+t\Omega_2(Q)-E'x+N't+h_2
                \},
              \label{BAf2}\\
\psi_3(Q)&=\frac{\theta\left(\int\limits_{\infty_1}^Qd\boldsymbol{v}+\boldsymbol{\Gamma}
-\boldsymbol{r}_3 \right)
                       \theta(\boldsymbol{D}-\boldsymbol{r}_3) }
              {\theta\left(\int\limits_{\infty_1}^Qd\boldsymbol{v}-\boldsymbol{D}
\right)
              \theta(\boldsymbol{\Gamma}) }
              \notag\\&\qquad\times
              \mathrm{exp}\{x\Omega_1(Q)+t\Omega_2(Q)
              +E'x-N't+h_3  \},
              \label{BAf3}
              \end{align}
where
 \begin{equation}
\boldsymbol{\Gamma}\boldsymbol{V}x+\boldsymbol{W}t-\boldsymbol{D}\label{boldgamma}
\end{equation}
and
\begin{equation}\boldsymbol{D}=\sum_{j=1}^g\int\limits_{\infty_1}^{Q_j}d\boldsymbol{
v}
-\boldsymbol{K}_{\infty_1}\label{boldd}
.\end{equation}
Here, $\boldsymbol{K}_{\infty_1}$ is the vector of Riemann constants with the
base point $\infty_1$. The constants $E,E',N,N'$ are defined by
$E=(E_1+E_2)/2$,
 $E'=-(E_1-E_2)/2$, and $N=(N_1+N_2)/2$,
 $N'=-(N_1-N_2)/2$,   where
 $E_1,E_2,N_1$ and $N_2$ are the basic constants from (\ref{omega1}) and
(\ref{omega2}).

The parameters appearing in the above expressions $\psi_i(Q,x,t)$
are defined in (\ref{windings}), and vectors $\boldsymbol{r}_{2,3}$
are defined in (\ref{rij}).

The proof of formulae (\ref{BAf1},\ref{BAf2},\ref{BAf3}) is based on
the standard arguments of the theory of algebro-geometric
integration: the Riemann theorem, which provides the condition {\bf
I.}, the non-speciality of the divisor $\mathcal{ D}$, which
guarantee the uniqueness of the function $\boldsymbol{\Psi}(Q,x,t)$,
and the periodicity properties (\ref{periodicity}) of the $\theta$ -
function, which ensure that the equations (\ref{BAf1}) -
(\ref{BAf3}) define a single-valued (meromorphic) function on $X$.
(For more details - see e.g. the similar proof for the usual,
one-component  NLS discussed in Ch. 4 of \cite{bbeim94}.)

We now fix some connected neighborhood $U$ of the point $z=\infty$
on $\mathbb{C}\mathbb{P}^1$ which has no branch points. Then, for
each $z\in U$, $\pi^{-1}(z)$ contains exactly three points denoted
by $Q_j\in X$, $j=1,2,3$, so that $Q_j\to\infty_j$ when
$z\to\infty$. For $z\in U$ the matrix function
\begin{equation}
\Psi(z,x,t)=\left(\boldsymbol{\Psi}(Q_1,x,t),
\boldsymbol{\Psi}(Q_2,x,t), \boldsymbol{\Psi}(Q_3,x,t) \right)
\label{psimatrix}\end{equation} is now correctly defined to enable
us to use the \cite{bbeim94} version of Krichever's method \cite{kr77} to
solve the VNSE.
This will require the asymptotic form for $\Psi(z,x,t)$, whose leading
term is

\begin{align*}
&\left(\begin{array}{lllll}
 \mathrm{e}^{\{-ixz-2\imath t
z^2\}}&\begin{array}{l}\vdots\\\vdots\\\vdots\end{array}
 &\begin{array}{l}
\frac{\theta(\boldsymbol{r}_2+\boldsymbol{\Gamma})\theta(\boldsymbol{D})}
                        {\theta(\boldsymbol{r}_2-\boldsymbol{D})\theta(\boldsymbol{\Gamma})}\\
 \times \mathrm{e}^{\{ixz+2\imath t
z^2\}}\\\times\mathrm{e}^{\{-E_1x+N_1t\}}\end{array}
 &\begin{array}{l}\vdots\\\vdots\\\vdots\end{array}
 &\begin{array}{l}\frac{\theta(\boldsymbol{r}_3+\boldsymbol{\Gamma})\theta(\boldsymbol{D})}
                        {\theta(\boldsymbol{r}_3-\boldsymbol{D})\theta(\boldsymbol{\Gamma})}\\
                         \times  \mathrm{e}^{\{ixz+2\imath t
z^2\}}\\\times\mathrm{e}^{\{-E_2x+N_2t\}}\end{array}\\
                         \dotfill
                         &\dotfill
                         &\dotfill
                         &\dotfill
                         &\dotfill\\
\begin{array}{l} \frac{\delta_2}{z}
\frac{\theta(\boldsymbol{r}_2-\boldsymbol{\Gamma})
  \theta(\boldsymbol{D}-\boldsymbol{r}_2)}
                        {\theta(\boldsymbol{D})\theta(\boldsymbol{\Gamma})}\\
                       \times\mathrm{e}^{\{-ixz-2\imath t
z^2\}}\\\times\mathrm{e}^{\{E_1x-N_1t\}}\end{array}
&\begin{array}{l}\vdots\\\vdots\\\vdots\end{array}
& \frac{z}{\delta_2} \mathrm{e}^{\{ixz+2\imath t z^2\}}
&\begin{array}{l}\vdots\\\vdots\\\vdots\end{array}
& \begin{array}{l}
\frac{\theta(\boldsymbol{r}_3-\boldsymbol{r}_2+\boldsymbol{\Gamma})
  \theta(\boldsymbol{D}-\boldsymbol{r}_2)}
                        {\theta(\boldsymbol{r}_3-\boldsymbol{D})\theta(\boldsymbol{\Gamma})}\\
 \times \mathrm{e}^{\{ixz+2\imath t
z^2\}}\\\times\mathrm{e}^{\{-(E_2-E_1)x+(N_2-N_1)t\}}
 \end{array}\\
\dotfill
&\dotfill
&\dotfill
&\dotfill
&\dotfill\\
 \begin{array}{l} \frac{\delta_3}{z}
\frac{\theta(\boldsymbol{r}_3-\boldsymbol{\Gamma})
  \theta(\boldsymbol{D}-\boldsymbol{r}_3)}
                        {\theta(\boldsymbol{D})\theta(\boldsymbol{\Gamma})}\\
 \times\mathrm{e}^{\{-ixz-2\imath t
z^2\}}\\\times\mathrm{e}^{\{E_2x-N_2t\}}\end{array}
 &\begin{array}{l}\vdots\\\vdots\\\vdots\end{array}
 &\begin{array}{l}
\frac{\theta(\boldsymbol{r}_2-\boldsymbol{r}_3+\boldsymbol{\Gamma})
  \theta(\boldsymbol{D}-\boldsymbol{r}_2)}
                        {\theta(\boldsymbol{r}_2-\boldsymbol{D})\theta(\boldsymbol{\Gamma})}\\
 \times \mathrm{e}^{\{ixz+2\imath t z^2\}}\\\times
 \mathrm{e}^{\{(E_2-E_1)x-(N_2-N_1)t\}}\end{array}
 &
 \begin{array}{l}\vdots\\\vdots\\\vdots\end{array}
 &\frac{z}{\delta_3} \mathrm{e}^{\{ixz+2\imath t z^2\}}
                            \end{array}\right)\\
 =&\left(
 \begin{array}{lllll}\qquad \qquad
 1 +
O\left(\frac{1}{z}\right)&\begin{array}{l}\vdots\\\vdots\\\vdots\end{array}
 &\begin{array}{l}\frac{\delta_2}{z}
\frac{\theta(\boldsymbol{r}_2+\boldsymbol
{\Gamma})\theta(\boldsymbol{D})}
                        {\theta(\boldsymbol{r}_2-\boldsymbol{D})\theta(\boldsymbol{\Gamma})}\\
 \times \mathrm{e}^{\{-E_1x+N_1t\}}\\+O\left(\frac{1}{z^2}\right)\end{array}
 &\begin{array}{l}\vdots\\\vdots\\\vdots\end{array}
 &\begin{array}{l}\frac{\delta_3}{z}\frac{\theta(\boldsymbol{r}_3+\boldsymbol{\Gamma})
 \theta(\boldsymbol{D})}
                        {\theta(\boldsymbol{r}_3-\boldsymbol{D})\theta(\boldsymbol{\Gamma})}\\
                         \times  \mathrm{e}^{\{-E_2x+N_2t\}}\\
+O\left(\frac{1}{z^2}\right)\end{array}\\
                         \dotfill
                         &\dotfill
                         &\dotfill
                         &\dotfill
                         &\dotfill\\
\begin{array}{l} \frac{\delta_2}{z}
\frac{\theta(\boldsymbol{r}_2-\boldsymbol{\Gamma})
  \theta(\boldsymbol{D}-\boldsymbol{r}_2)}
                        {\theta(\boldsymbol{D})\theta(\boldsymbol{\Gamma})}\\
                       \times\mathrm{e}^{\{E_1x-N_1t\}}\\+O\left(\frac{1}{z^2}\right)\end{array}
&\begin{array}{l}\vdots\\\vdots\\\vdots\end{array}
& \qquad \qquad 1 + O\left(\frac{1}{z}\right)
&\begin{array}{l}\vdots\\\vdots\\\vdots\end{array}
& \begin{array}{l} \frac{\delta_3}{z}
\frac{\theta(\boldsymbol{r}_3-\boldsymbol{r}_2+\boldsymbol{\Gamma})
  \theta(\boldsymbol{D}-\boldsymbol{r}_2)}
                        {\theta(\boldsymbol{r}_3-\boldsymbol{D})\theta(\boldsymbol{\Gamma})}\\
 \times\mathrm{e}^{\{-(E_2-E_1)x+(N_2-N_1)t\}}\\+O\left(\frac{1}{z^2}\right)
 \end{array}\\
\dotfill
&\dotfill
&\dotfill
&\dotfill
&\dotfill\\
 \begin{array}{l} \frac{\delta_3}{z}
\frac{\theta(\boldsymbol{r}_3-\boldsymbol{\Gamma})
  \theta(\boldsymbol{D}-\boldsymbol{r}_3)}
                        {\theta(\boldsymbol{D})\theta(\boldsymbol{\Gamma})}\\
 \times\mathrm{e}^{\{E_2x-N_2t\}}\\+O\left(\frac{1}{z^2}\right)\end{array}
 &\begin{array}{l}\vdots\\\vdots\\\vdots\end{array}
 &\begin{array}{l}\frac{\delta_2}{z}
\frac{\theta(\boldsymbol{r}_2-\boldsymbol{r}_3+\boldsymbol{\Gamma})
  \theta(\boldsymbol{D}-\boldsymbol{r}_2)}
                        {\theta(\boldsymbol{r}_2-\boldsymbol{D})\theta(\boldsymbol{\Gamma})}\\
 \times
 \mathrm{e}^{\{(E_2-E_1)x-(N_2-N_1)t\}}\\+O\left(\frac{1}{z^2}\right)\end{array}
 &
 \begin{array}{l}\vdots\\\vdots\\\vdots\end{array}
 &\qquad \qquad 1 + O\left(\frac{1}{z}\right) \end{array}\right)\\
 \\\\
&\qquad \qquad\qquad \qquad \times \mathrm{exp}\{ \imath
z x J +2\imath z^2t J
\}
\left(\begin{array}{ccc}
  1&{}&{}\\{} &\frac{z}{\delta_2}&{}\\{}&{}&\frac{z}{\delta_3}\end{array}
\right) ,
 \end{align*}
where
\begin{equation} J=\left(\begin{array}{cc}-1&\boldsymbol{0}^T\\
                            \boldsymbol{0}&1_2  \end{array}\right).
\label{jmatrix}
                             \end{equation}

\begin{theorem}\label{maintheorem}
 Let ${\mathcal D}
=Q_1+\ldots,Q_g$ be non-special divisor of degree $g$ satisfying
the reality condition,
\begin{equation}\label{divsym}
\overline{\boldsymbol{D}}=\boldsymbol{D}, \quad
\boldsymbol{D}\equiv \sum_{j=1}^g\int\limits_{\infty_1}^{Q_j}d\boldsymbol{ v}
-\boldsymbol{K}_{\infty_1}
\end{equation}
Then the solution of the Manakov system reads

\begin{align}
q_{1,2}(x,t) &=2\imath A_{1,2}\frac{\theta
\left( \boldsymbol{V}x+
\boldsymbol{W}t-\boldsymbol{D}+\boldsymbol{r}_{2,3}
\right) }{\theta \left( \boldsymbol{V}x+\boldsymbol{W}
t - \boldsymbol{D}\right) }\mathrm{exp}\left\{
-E_{1,2}x+N_{1,2}t\right\}\notag
, \\
A_{1,2} &=\delta_{2,3}\mathrm{exp}
\left\{ \imath\;\mathrm{arg}\left(
\frac{\theta(\boldsymbol{D})}{\theta(\boldsymbol{r}_{2,3}-\boldsymbol{D})}
 \right)\right\}.\label{solution}
\end{align}
The constants
$E_{1,2}, N_{1,2}$ and $\delta_{2,3}$    are
defined in (\ref{e12}), (\ref{n12}) and (\ref{delta23}),
the vectors $\boldsymbol{r}_{2,3}$ are defined in (\ref{rij}),
and the winding vectors $\boldsymbol{V}$ and
$\boldsymbol{W}$ are given in (\ref{wind}) - (\ref{windings}).
\end{theorem}

Following the methodology of \cite{bbeim94}, we shall first prove
two general lemmas.
\begin{lemma}
Let $\Psi(z,x,t)$ be $3\times3$ matrix function holomorphic in some
neighborhood of infinity on the Riemann sphere smoothly dependent on
$x,t$ with the following asymptotic expansion at infinity

\begin{align}
\left.\Psi(z,x,t)\right|_{z\rightarrow\infty}
&=\left[1_3 +\sum\limits_{k=1}^{\infty} m_k(x,t) z^{-k} \right]\notag\\
&\times \exp\{\imath z x J+2\imath z^2 t J\}C(z),\label{AI1}
\end{align}
where $J$ is defined in (\ref{jmatrix}) and $C_x(z)=C_t(z)=0$
Then assuming that (\ref{AI1}) is differentiable in $x$ and $t$

\begin{eqnarray}
\Psi_x\Psi^{-1}&=&M(z)+o(z^{-1}),\notag
\label{AI2}\\
\Psi_t\Psi^{-1}&=&B(z)+o(z^{-1}),\notag
\end{eqnarray}
 where $z\longrightarrow\infty$ and
\begin{eqnarray}
M(z)=\imath z J+\imath [m_1,J]\equiv \left(\begin{array}{cc}
-\imath z&\boldsymbol{q}^T\\
 -\boldsymbol{p}&\imath z 1_2      \end{array}\right),\label{AI3}
\end{eqnarray}
where
\[\boldsymbol{p}=2\imath\left(\begin{array}{c}m_{1,21}
   \\m_{1,31}\end{array}\right) ,\qquad
 \boldsymbol{q}=2\imath\left(\begin{array}{c}m_{1,12}
   \\m_{1,13}\end{array}\right)   \]
and

$$
B(z)=2\imath z^2 J +2\imath z [m_1,J]
$$
\begin{eqnarray}
+2\imath [m_2,J]-2\imath [m_1,J] m_1.
\label{AI4}\end{eqnarray}

\label{lem1}
\end{lemma}
\begin{proof} Direct calculations
\end{proof}

\begin{lemma}\label{lem2}
Suppose that $\Psi(z)$ satisfies the condition of the Lemma \ref{lem1}
and equations,
\[ \Psi_{x}=M\Psi,\quad  \Psi_{t}=B\Psi  \]
with  $M(z)$ and $B(z)$ defined in (\ref{AI3}) and (\ref{AI4}). Then $M(z)$
and $B(z)$ are of the form
 presented in the Introduction, but $\boldsymbol{p}$ replaces
$\overline{\boldsymbol{q}}$.
  \end{lemma}

\begin{proof}
>From $\Psi_x=M\Psi $ it follows that
\begin{equation}
m_{1,x}=-\imath[m_2,J]+\imath[m_1,J]m_1.\label{AI5}
\end{equation}
Given any $3\times3$ matrix $A$, we can represent it as

\[ A=\left(\begin{array}{ccc}a_{11}&0&0\\
                             0&a_{22}&a_{23}\\
                             0&a_{32}&a_{33}  \end{array} \right)
+\left(\begin{array}{ccc}a&a_{12}&a_{13}\\
                             a_{21}&0&0\\
                             a_{31}&0&0 \end{array} \right)\equiv
A_{d}+A_{off}.\]
Note that $[A,J]_d=0$. To prove the Lemma we only need to check that

\[B_0\equiv 2\imath [m_2,J]-2\imath[m_1,J]m_1=\left(\begin{array}{cc}
\imath\boldsymbol{p}^T\boldsymbol{q}&\imath\boldsymbol{q}_x^T\\
  \imath\boldsymbol{p}_x&-\imath\boldsymbol{p}\boldsymbol{q}^T
 \end{array}\right).\]
Direct calculation shows that

\[(B_0)_d
=\left(-2\imath[m_1,J_1]m_1\right)_d=\left(\begin{array}{cc}\imath\boldsymbol{p}^T
\boldsymbol{q}&\boldsymbol{0}^T\\
\boldsymbol{0}&-\imath\boldsymbol{p}\boldsymbol{q}^T\end{array}
\right).\]
At the same time taking into the account (\ref{AI5})

\[(B_0)_{off} =-2(m_{1,x})_{off}=\left(\begin{array}{cc}
0&\imath\boldsymbol{q}_x^T\\
\imath\boldsymbol{p}_x&0_2\end{array}\right). \]

\end{proof}

We can now proceed with the proof of theorem \ref{maintheorem}.
Consider the matrix $\Psi(z,x,t)$ defined in (\ref{psimatrix});  we claim
that

{\bf (A)} $\Psi(z,x,t)$ satisfies conditions of the Lemma \ref{lem1} with
\[ C(z)=\left(\begin{array}{ccc}1&0&0\\
                                0&\frac{z}{\delta_2}&0 \\
                                0&0&\frac{z}{\delta_3}
\end{array}\right), \]

{\bf (B)} $\Psi(z,x,t)$ satisfies conditions of the Lemma
\ref{lem2}
\begin{proof}
{\bf (A)} has already been established - see the asymptotic form
of $\Psi(z,x,t)$ presented above. Moreover, we have from this form the
following
expressions for the relevant vectors $\boldsymbol{q}$ and $\boldsymbol{p}$.

\begin{align}
q_{1,2} &=2\imath \delta_{2,3}\frac{\theta
\left( \boldsymbol{r}_{2,3} + \boldsymbol{\Gamma}
\right)\theta\left(\boldsymbol{D}\right)}{\theta \left(
\boldsymbol{r}_{2,3} - \boldsymbol{D}\right)\theta\left(\boldsymbol{\Gamma}
\right)}\mathrm{exp}\left\{ -E_{1,2}x+N_{1,2}t\right\}\label{q0} , \\\notag\\
p_{1,2} &=2\imath \delta_{2,3}\frac{\theta
\left( \boldsymbol{r}_{2,3} - \boldsymbol{\Gamma}
\right)\theta\left(\boldsymbol{r}_{2,3} - \boldsymbol{D}\right)}{\theta
\left(
 \boldsymbol{D}\right)\theta\left(\boldsymbol{\Gamma}
\right)}\mathrm{exp}\left\{ E_{1,2}x-N_{1,2}t\right\}.\label{p0}
\end{align}

To prove {\bf (B)} it is enough to show that
$\boldsymbol{\Psi}_x(Q)=M(z)\boldsymbol{\Psi}(Q)$ and
$\boldsymbol{\Psi}_t(Q)=B(z)\boldsymbol{\Psi}(Q)$ for all $Q \in X$.
Consider the first equation. Put

\[ \boldsymbol{f}(Q)=\boldsymbol{\Psi}_x(Q)-M(z)
\boldsymbol{\Psi}(Q) \]
and note that if $Q$ is in a neighborhood of the point
$\infty_{j}$, then
\[ \boldsymbol{f}(Q)\equiv \left({F}(z)\right)_j,\qquad
 F(z)=\Psi_x(z)-M(z)\Psi(z),\quad \pi(Q) = z, \]
where $(A)_{j}$ denotes the $j$th column of a  matrix $A$.
We have at $z\rightarrow \infty$

\begin{eqnarray}
F(z)&=&[\Psi_x\Psi^{-1}-M]\Psi\\
&=&
O\left(\frac{1}{z}\right)\left(\begin{array}{ccc}
e&O(1)e^{-1}&O(1)e^{-1}\\
O(z^{-1})e&O(z) e^{-1}& O(1)e^{-1}\\
O(z^{-1})e&O(1)e^{-1}&O(z) e^{-1}
\end{array}\right),\notag
\end{eqnarray}
where $e=\mathrm{exp}(-\imath z x-2\imath z^2 t)$.

It follows that
\begin{eqnarray*}
\boldsymbol{f}(Q)=\begin{cases}
o(1)\exp\{-\imath z t -2\imath z^2 x
\}&\text{at}\quad Q\rightarrow \infty_{1},\\
O(1)\exp\{\imath z t +2\imath z^2 x
\}&\text{at}\quad Q\rightarrow \infty_{2,3}.\end{cases}
\end{eqnarray*}

Hence by the non-speciality of the divisor ${\mathcal D}$
we conclude that $ \boldsymbol{f}(Q)\equiv 0 $ (cf. \cite{kr77}; see also
Corollary 2.26 \cite{bbeim94}),
which implies $\boldsymbol{\Psi}_x(Q)=M(z)
\boldsymbol{\Psi}(Q)$ and the validity
of the first equation follows.

The second equation can be proven by considering
\[ \widetilde{\boldsymbol{f}}(Q)=\boldsymbol{\Psi}_t(Q)
-B(z)\boldsymbol{\Psi}(Q) \]
and applying exactly the same arguments.

\end{proof}

As an immediate consequence we arrive to the following corollary.

\begin{cor} The functions $\boldsymbol{q}(x,t)$ and $\boldsymbol{p}(x,t)$
defined in (\ref{q0}) and (\ref{p0}) form a solution of the equations

\begin{align*}
&\imath
\boldsymbol{q}_t+\boldsymbol{q}_{xx}+2\boldsymbol{p}^T
\boldsymbol{q}\boldsymbol{q}=0,\\
-&\imath \boldsymbol{p}_t+\boldsymbol{p}_{xx}+2\boldsymbol{q}^T
\boldsymbol{p}\boldsymbol{p}=0.
\end{align*}
\end{cor}

Using conjugation properties (\ref{VWcon}), (\ref{rcon}), (\ref{ENcon}),
(\ref{deltasym})  of the quantities $\boldsymbol{V}$,
$\boldsymbol{W}$, $\boldsymbol{r}_{2,3}$, $E_{1,2}$, $N_{1,2}$,
$\delta_{2,3}$,
the symmetry property   (\ref{thetasym2}) of the $\theta$-function, and
taking into account
condition (\ref{divsym}) which we imposed on the divisor ${\mathcal D}$ we
see that

\begin{align}
{p}_{j}(x,t)={\alpha_{j}}\bar{q}_{j}(x,t),\quad j = 1, 2,
\end{align}
where

\begin{align}
\alpha_{1,2}=\frac{\theta(\boldsymbol{r}_{2,3}-\boldsymbol{D})
\theta(\boldsymbol{r}_{2,3}+\boldsymbol{D})}{\theta^2(\boldsymbol{D})}
=\left|\frac{\theta(\boldsymbol{r}_{2,3}-\boldsymbol{D})}
{\theta(\boldsymbol{D})}\right|^2.
\end{align}
Hence $q_1$ and $q_2$ satisfy the evolution equations

\begin{align*}
\imath\frac{\partial q_1}{\partial t}+
\frac{\partial^2 q_1}{\partial x^2}+2(\alpha_1|q_1|^2+\alpha_2|q_2|^2)q_1=0,
\\
\imath\frac{\partial q_2}{\partial t}+
\frac{\partial^2 q_2}{\partial x^2}+2(\alpha_1|q_1|^2+\alpha_2|q_2|^2)q_2=0.
\end{align*}

A trivial rescaling

\begin{equation}q_1 \mapsto \sqrt{\alpha_1}{q}_1, \quad
q_2 \mapsto \sqrt{\alpha_2}{q}_2
 \end{equation}
complete the proof of the theorem.

We emphasize that the solution (\ref{solution}) obtained is effective
because computation of all parameters of solution, such as winding
vectors, constants coming into exponentials was reduced to
computation of holomorphic differentials, their periods and
$\theta$-functions. These last computations are well algorithmized
by Deconinck and van Hoeij in Maple for arbitrary curve, see also
their paper \cite{dh01}. At the end of the paper we describe
a computing procedure based on Maple software to compute
algebro-geometric solutions to the Manakov system.

\section{Example: solution in elliptic functions }
In this section we shall show how the construction works in the
simplest case of genus one. The spectral curve $X$ reads
\begin{equation}
(w+2\imath z^2)(w-2\imath z^2)^2
+(2\lambda_2z+\lambda_3)(w-2\imath z^2)+\mu_0=0,\label{gen1}
\end{equation}
where the parameters $\lambda_2,\lambda_3$ and $\mu_0$ are
\begin{align}
\lambda_2&= \alpha_3=\imath
(\boldsymbol{\gamma}_1^T\boldsymbol{\beta}_0
+\boldsymbol{\gamma}_0^T\boldsymbol{\beta}_1)
=-\imath( \boldsymbol{q}^T\bar{\boldsymbol{q}}_x-
\boldsymbol{q}^T_x\bar{\boldsymbol{q}}),  \notag\\
\lambda_3&=\alpha_4- \imath\boldsymbol{\gamma}_2^T\boldsymbol{\beta}_0-
\imath\boldsymbol{\gamma}_0^T\boldsymbol{\beta}_2
=\boldsymbol{q}_x^T\bar{\boldsymbol{q}}_x+(\boldsymbol{q}^T
\bar{\boldsymbol{q}})^2
,\label{abc}\\
\mu_0&=\imath\vert q_{1x}q_2-q_{2x}q_1\vert^2.
\notag\end{align}
As commented earlier, the expression for $\mu_0$, and the fact that the last
term in (\ref{gen1}) is a polynomial of degree zero is obtained from
Maple. We also remark that parameters $\lambda_{2,3}$ are real whilst
$\mu_0$ be pure imaginary as that was stated in (\ref{lmbar})

Let us denote quantity (see (\ref{leadpow}))
$\Delta=\lambda_2^2-4\imath \mu_0>0$.
The discriminant of the curve has not multiple roots
if and only if
\begin{align*}\mu_0 \Delta(27\Delta^2-64\lambda_3^3)[
-27\imath\mu_0(\Delta+\imath\mu_0)^3+\lambda_3^3(\lambda_3^3
+54\imath\mu_0\Delta-270\mu_0)]\neq 0
\end{align*}

\begin{figure}[ht]
\includegraphics[width=0.8\textwidth]{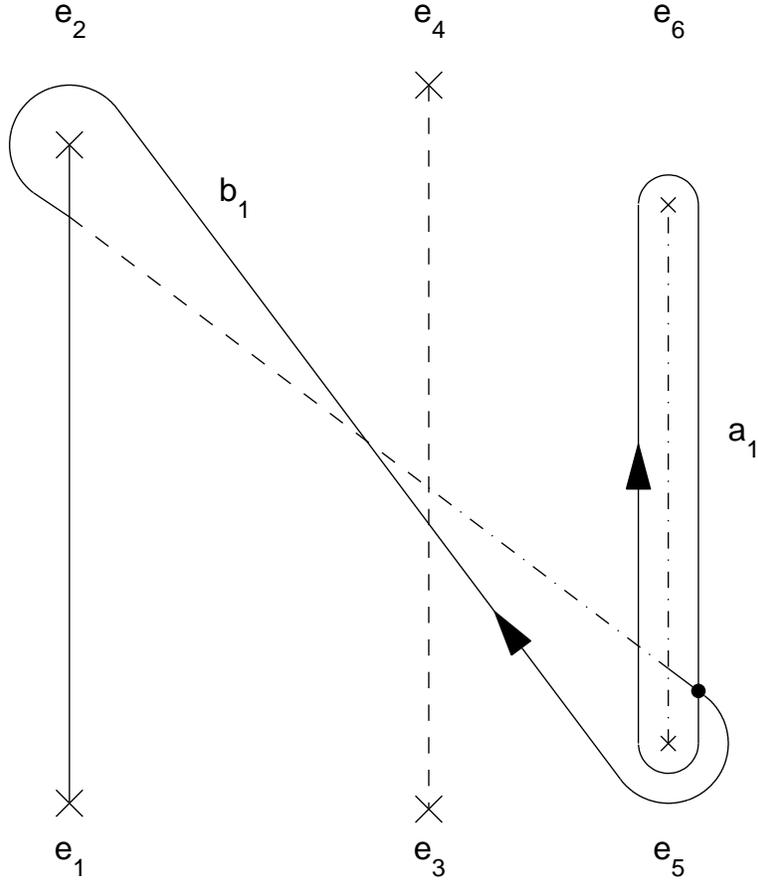}
\caption{Basis of cycles of the elliptic curve $X$ of  genus $1$.
The solid line, dashed line and  dashed-dotted
lines denote paths on the first, second and third sheets correspondingly.
The cuts between the first and second, second and the third and third and
the first sheets are denoted as correspondingly}
\label{fig1}
\end{figure}

This curve is of genus 1 and admits the following behavior on the
sheets at infinities
\begin{align}\begin{split}
z=\frac{1}{\xi},\quad&
w=-\frac{2\imath}{\xi^2}-\frac{\imath}{2}\lambda_2\xi+O(\xi^2)\quad\text{on
the first sheet},\\
z=\frac{1}{\xi},\quad&
w=\frac{2\imath}{\xi^2}+\frac{\imath}{4}(\lambda_2+\sqrt{\Delta})
\xi+O(\xi^2)\quad\text{on the second sheet},\\
z=\frac{1}{\xi},\quad&
w=\frac{2\imath}{\xi^2}+\frac{\imath}{4}(\lambda_2-\sqrt{\Delta})
\xi+O(\xi^2)\quad\text{on the third sheet}.\end{split}\label{sheets1}
\end{align}

The holomorphic differential on $X$ be of the form
\begin{eqnarray}
du=\frac{(w-2\imath z^2)dz}{3w^2+4\imath w z^2
+4z^4 +\lambda_2z+ \lambda_3}.\label{holdiff1}
\end{eqnarray}
Denote $ A=\oint_{\mathfrak{a}} du$ its $\mathfrak{a}$-period, then
the normalized holomorphic differential $dv=du/A$.

In the case considered we have $r_2=-r_3=r$,
\begin{equation}  r=\frac{1}{A}\int\limits_{\infty_1}^{\infty_2}du.
\label{ellipticr}\end{equation}
Indeed, because of expansions (\ref{sheets}) the function on the curve
$w-2\imath z^2$ has second order pole at the first sheet and first
order zeros on the second and third sheets. Then the Abel theorem says
$r_2+r_3=0$.

The auxiliary winding numbers (\ref{windings})
are computed with the aid of (\ref{sheets}) as follows
\begin{align}\begin{split} V^{(1)}&=\frac{1}{4A},\quad
              V^{(2)}=\frac{-\sqrt{\Delta}-\lambda_2
              }{8A\sqrt{\Delta}},\quad
              V^{(3)}=\frac{-\sqrt{\Delta}+\lambda_2 }
                           {8A\sqrt{\Delta}}.\\
W^{(1)}&=0,\quad W^{(2)}=\frac{\imath \lambda_3\mu_0 }{8A
 \sqrt{\Delta^3} },\quad W^{(2)}=-\frac{\imath \lambda_3\mu_0 }
{8A\sqrt{\Delta^3}. }\end{split}\label{elwind}
\end{align}
Therefore we have for the main winding numbers
\begin{equation} V=\frac{1}{2A}, \quad  W=0\label{mainwind}.
\end{equation}

To perform further calculations it is convenient to transform the curve
$X$ to
the form of standard Weierstrass cubic.
Namely there exists a  birational transformation $T$
\begin{align}\begin{split}
T&:X\rightarrow\widetilde{X}\quad    z=\frac{y-8\lambda_2}{8x},\quad
w=\frac{\imath}{32x^2}\left((y^2-8\lambda_2)^2-8x^3\right)\\
T^{-1}&:\widetilde{X}\rightarrow {X}\quad x=4\imath(w-2\imath z^2),\quad
y=32\imath z (w-2\imath z^2)+8\lambda_2
\end{split}\label{birat}\end{align}
of the curve $X$ to   $\widetilde{X}=(x,y)$
\begin{equation}y^2=4x^3-g_2x-g_3\label{wcubic}\end{equation}
with parameters
\begin{equation}
g_2 = 64\lambda_3,\quad  g_3 = -64\Delta\label{g2g3}.
\end{equation}
The discriminant of $\widetilde{X}$, coincides with
multiplier  $27\Delta^2-64\lambda_3^3$ of the expression for the
discriminant $X$. In what follows we shall use Weierstrass functions
of the curve $\widetilde{X}$.

The transformation $T$ maps canonical holomorphic differential $du$ to
$dx/y$. The point
$2\omega r$ appeared to be a zero of the Weierstrass $\wp$-function,
\begin{equation} \wp(2\omega r)=0\label{wpzero}
\end{equation}
what follows from the fact that the map $T$ is mapping $\infty_1$ of
the curve $X$ to the $\infty$ of the curve $\widetilde{X}$ and
  $\infty_{2,3}$ of the curve $X$ to points $(0,\pm\imath \sqrt{g_3})$
  of  curve $\widetilde{X}$. Remark that solution of (\ref{wpzero}) are
given in \cite{ez82} in terms of Eisenstein series what will be of
importance for developing computational algorithms.

Direct computations gives\footnote{We shall give this computation for
  completeness in the Appendix}
\begin{equation}
E_1=-E_2=E=-\frac12(\zeta(2\omega r)-2\eta r)
,\quad N_{1,2}=0 .\label{en}
\end{equation}

The equalities $W=0, N_{1,2}=0$ are in accordance to our
analysis which leads to the
statement that the time $t=t_2$ is stationary for the flow yielded
the Manakov matrix $B$ (see \ref{zerocurv}).

The solution has the form
\begin{align}
\begin{split}
q_{1,2}(x)&=2\imath A_{1,2}\frac{\vartheta_3(Vx\pm r -D
    )}{\vartheta_3(Vx -D)} \mathrm{e}^{\pm E x},\\
A_{1,2}&=\pm\frac{\imath}{4\sqrt{2}A}\frac{\lambda_2\mp\sqrt{\Delta}}
{\sqrt{\Delta}}\frac{\vartheta_1'(0)}{\vartheta_1(r)}
\mathrm{exp}\left\{ \imath\mathrm{arg}\left(
    \frac{\vartheta_3(D)}{\vartheta_3(\pm r -D)} \right)  \right\},
\end{split}\label{elsol}\end{align}
where $\theta(z)=\vartheta_3(z)$ is canonical $\theta$-function
of the genus one curve (\ref{gen1}), $D$ is arbitrary constant satisfying
$\theta(D)\neq 0$. Substitution of the elliptic solution (\ref{elsol})
to the expressions for levels of the integrals of motion (\ref{abc})
leads to equivalences and permit moreover to compute the link between periods
$A$ and $2\omega$ as\footnote{see Appendix for details}
\begin{equation} A=-2\omega.\label{aomega}  \end{equation}

It is remarkable that the simplest solution to the Manakov system,
i.e. solution of genus one independent in time and as the result it
soliton limit enable to yield Manakov soliton (\ref{manakov}). We
can guess (\ref{manakov}) can be obtain as the result of
degeneration of the genus three curve.

\section{Summary: computational algorithm }
This section is addressed to a reader who wants to know the computing
algorithm without going through the arguments of the paper.
The procedure to compute algebro-geometric solutions to
the Vector Nonlinear Schr\"{o}dinger equation
\begin{eqnarray*}
\imath \frac{\partial q_{1}}{\partial t}
+\frac{\partial ^{2}q_{1}}{\partial x^{2}}
+2\left( |q_{1}|^{2}+|q_{2}|^{2}\right) q_{1} &=&0, \\
\imath \frac{\partial q_{2}}{\partial t}
+\frac{\partial ^{2}q_{2}}{\partial x^{2}}
+2\left( |q_{1}|^{2}+|q_{2}|^{2}\right) q_{2} &=&0
\end{eqnarray*}
can be formulate as follows.
\begin{itemize}
\item Fix positive integer $n\in \{2,3,\ldots\}$.
\item Fix a polynomial in two variables (algebraic curve)
\begin{align*}
f(z,w)=(w+\frac{\imath}{2} (2z)^n)
(w-\frac{\imath}{2} (2z)^n)^2
+(w-\frac{\imath}{2} (2z)^n)P_{n-1}(z)
+\imath P_{n-2}(z)=0,
\end{align*}
with arbitrary real polynomials $P_{n-2}$ and $P_{n-1}$ of degrees
$n-2$ and $n-1$ correspondingly. Compute its genus $g$  by using
\cite{dh01} and Maple. For
polynomials   $P_{n-2}$ and $P_{n-1}$ in general position $g=2n-3$.

\item Compute the vector of  holomorphic differentials
$d\boldsymbol{u}(Q)=(du_1(Q),\ldots, du_g(Q))^T $.
\begin{align*}
du_j(Q)&=\frac{\imath z^j}{\frac{\partial}{\partial w} f(z,w) }dz,\quad
j=0,\ldots,n-3,\\
du_j(Q)&=\frac{z^{n-2-j} \left(w-\frac{\imath}{2}(2z)^n\right)
  }{\frac{\partial}{\partial w} f(z,w)}\,dz,\quad j=n-2,\ldots,2n-3.
\end{align*}
\item Compute vector of normalized holomorphic differentials
 \[ d\boldsymbol{v}(Q)=A^{-1} d\boldsymbol{u}(Q)  \]
\item Compute auxiliary winding vectors  $\boldsymbol{V}^{(i)},
\boldsymbol{W}^{(i)},\boldsymbol{Z}^{(i)}$, $i=1,2,3$ from expansions
\[
\left.\int\limits_Q^Pd\boldsymbol{v}\right|_{P\to\infty_i}
=O(1)+\boldsymbol{V}^{(i)}\xi+
\boldsymbol{W}^{(i)}\xi^2+\boldsymbol{Z}^{(i)}\xi^3+\ldots,\quad i=1,2,3,
 \]
Set for main winding vectors
\begin{align*} \boldsymbol{V}&=\imath\boldsymbol{V}^{(1)}
-\imath \boldsymbol{V}^{(2)}-\imath \boldsymbol{V}^{(3)},\quad
 \boldsymbol{W}=4\imath \boldsymbol{W}^{(1)}
-4\imath \boldsymbol{W}^{(2)}-4\imath \boldsymbol{W}^{(3)}
.\end{align*}
\item Compute vectors $\boldsymbol{r}_{2,3}$
\[\boldsymbol{r}_{2,3}=\int\limits_{\infty_1}^{\infty_{2,3}}d\boldsymbol{v}
.\]
\item Compute  6 constants
\begin{align*}
E_{1,2}&=\imath
\left(\partial_{\boldsymbol{V}^{(1)}}-
 \partial_{\boldsymbol{V}^{(2,3)}}\right)
\mathrm{ln}\theta[\varepsilon](\boldsymbol{r}_{2,3})-
\imath\partial_{\boldsymbol{V}^{(3,2)}}
\mathrm{ln}\left[ { \theta[\varepsilon](\boldsymbol{r}_{3,2})}
{\theta[\varepsilon](\boldsymbol{r}_{2,3}-\boldsymbol{r}_{3,2} )}
 \right]\\&-\imath c_1^{(2,3)}- \imath c_1^{(1)},\\
N_{1,2}&=4\imath
\left(\partial_{\boldsymbol{W}^{(1)}}-
 \partial_{\boldsymbol{W}^{(2,3)}}\right)
\mathrm{ln}\theta[\varepsilon](\boldsymbol{r}_{2,3})-
4\imath\partial_{\boldsymbol{W}^{(3,2)}}
\mathrm{ln}\left[ { \theta[\varepsilon](\boldsymbol{r}_{3,2})}
{\theta[\varepsilon](\boldsymbol{r}_{2,3}-\boldsymbol{r}_{3,2} )}
 \right]\\
&-2\imath\left( \partial^2_{\boldsymbol{V}^{(1)},\boldsymbol{V}^{(2,3)}}-
 \partial^2_{\boldsymbol{V}^{(2,3)}, \boldsymbol{V}^{(1)}}\right)
\mathrm{ln}\theta[\varepsilon](\boldsymbol{r}_{2,3})   \\
&+2\imath\partial^2_{\boldsymbol{V}^{(3,2)},\boldsymbol{V}^{(2,3)}}
 \mathrm{ln}
\left[\frac{\theta[\varepsilon](\boldsymbol{r}_{3,2})}
{\theta[\varepsilon]
(\boldsymbol{r}_{2,3}-\boldsymbol{r}_{3,2})}\right]-\imath  c_2^{(2,3)}-
\imath  c_2^{(1)}
, \notag\\
\delta_{2,3}&=\frac{\imath}
{\theta[\varepsilon](\boldsymbol{r}_{2,3})}\sqrt{
\partial_{\boldsymbol{V}^{(2,3)}}
\theta[\varepsilon](\boldsymbol{0})}\sqrt{
\partial_{\boldsymbol{V}^{(1)}}
\theta[\varepsilon](\boldsymbol{0})}
,\end{align*}
where $[\varepsilon]$ be non-singular odd characteristic,
$\partial_{\boldsymbol{V}}$, $\partial_{\boldsymbol{W}}$ and
$\partial_{\boldsymbol{Z}}$ are
directional derivatives,
\begin{align*}
\partial_{\boldsymbol{V}}=&\sum_{k=1}^gV_k\frac{\partial}{\partial v_k},\quad
\partial^2_{\boldsymbol{V},\boldsymbol{W}}=\sum_{k=1}^g\sum_{l=1}^gV_kW_l
\frac{\partial^2}{\partial v_k\partial v_l},\quad \text{etc.}
\end{align*}
and also 6 constants, at $i=1,2,3$,
\begin{align*}
c_1^{(i)}&=\frac{1}{2}\frac{\partial^{2}_{\boldsymbol{V}^{(i)},\boldsymbol{V}^{(i)}}
\theta[\varepsilon](\boldsymbol{0}) }
{\partial_{\boldsymbol{V}^{(i)}}
  \theta[\varepsilon](\boldsymbol{0})}
-\frac{\partial_{\boldsymbol{W}^{(i)}} \theta[\varepsilon](\boldsymbol{0}) }
{\partial_{\boldsymbol{V}^{(i)}}
  \theta[\varepsilon](\boldsymbol{0})},\\
c_2^{(i)}&=-\left(\frac{\partial_{\boldsymbol{W}^{(i)}}
\theta[\varepsilon](\boldsymbol{0}) }
{\partial_{\boldsymbol{V}^{(i)}}
\theta[\varepsilon](\boldsymbol{0})}\right)^2
+\frac{1}{3}\frac{\partial^3_{\boldsymbol{V}^{(i)},\boldsymbol{V}^{(i)},\boldsymbol{V}^{(i)}
}
 \theta[\varepsilon](\boldsymbol{0}) }
{\partial_{\boldsymbol{V}^{(i)}} \theta[\varepsilon](\boldsymbol{0})}
+\frac{\partial_{\boldsymbol{Z}^{(i)}} \theta[\varepsilon](\boldsymbol{0}) }
{\partial_{\boldsymbol{V}^{(i)}} \theta[\varepsilon](\boldsymbol{0})}
.\end{align*}

\item The algebro-geometric solution is of the form
\begin{align*}
q_{1,2}(x,t) &=2\imath A_{1,2}\frac{\theta
\left( \boldsymbol{V}x+
\boldsymbol{W}t-\boldsymbol{D}+\boldsymbol{r}_{2,3}
\right) }{\theta \left( \boldsymbol{V}x+\boldsymbol{W}
t - \boldsymbol{D} \right) }\mathrm{exp}\left\{ -E_{1,2}x+N_{1,2}t\right\}
, \label{solution1}
 \\
A_{1,2} &=\delta_{2,3}\mathrm{exp}
\left\{ \imath\;\mathrm{arg}\left(
\frac{\theta(\boldsymbol{D})}{\theta(\boldsymbol{r}_{2,3}-\boldsymbol{D})}
 \right)\right\},
\end{align*}
where $\boldsymbol{D}$ is arbitrary vector satisfying condition
$\theta( \boldsymbol{D})\neq 0$.
 \end{itemize}

\section{Appendix} Write in addition to (\ref{elwind})
the third auxiliary winding numbers
\begin{equation}Z^{(1)}=0,\quad Z^{(2,3)}=\pm \frac{\imath}{16}
  \frac{\lambda_2\mu_0\lambda_3^2}{A\Delta^{\frac52}}
\end{equation}
The constants $c_l^{(k)},$ $l=1,2$ and   $k=1,2,3$ are
\begin{equation}c_1^{(1)}== 0,
\quad  c_1^{(2,3)}==\frac{\lambda_3(\lambda_2\pm\sqrt{\Delta})}{4\Delta}
\end{equation}
and
\begin{align*}
c_2^{(1)}&=-\frac{1}{48
  A^2}\frac{\vartheta_1'''(0)}{\vartheta_1'(0)},\\
c_2^{(2,3)}&=\frac{\imath \mu_0\lambda_3^2 (\imath \mu_0-\lambda_2^2\pm
\sqrt{\Delta}) }{(\lambda_2\mp\sqrt{\Delta})^2\Delta^2} -\frac{1}{192
  A^2} \frac{(\lambda_2\pm\sqrt{\Delta})^2}{\Delta^2}
\frac{\vartheta_1'''(0)}{\vartheta_1'(0)}.
\end{align*}

Compute first $E_1-E_2$. We have
\begin{align*}
E_1-E_2&=\imath (X_{12}-X_{13})-\imath (X_{32}-X_{23})+\imath
(c_1^{(3)}-c_1^{(2)})\\
&=2\imath V^{(1)}\left(\mathrm{ln}\, \vartheta_1(r) \right)' -
\imath ( V^{(2)} + V^{(3)}  ) \left(\mathrm{ln}\, \vartheta_1(2r) \right)'
-\frac{\imath \lambda_3 }{2\sqrt{\Delta}}\\
&=\frac{\imath}{4A}[4\omega\zeta(2\omega r )-8\eta\omega r]
+\frac{\imath}{4A}[2\omega\zeta(4\omega r )-8\eta\omega r]
-\frac{\imath \lambda_3 }{2\sqrt{\Delta}}.
\end{align*}
Apply the duplication formula
\[ \zeta(2z)=2\zeta(z)+\frac12\frac{\wp''(z)}{\wp'(z)},\]
which in the case considered reads
\[ \zeta(4\omega r)=\zeta(2\omega r)
-\frac{2\lambda_3}{\sqrt{\Delta}}
  \]
to obtain
\begin{equation}E_1-E_2=V(4\omega\zeta(2\omega r) -8\eta\omega
r)-\frac{\imath\lambda_3}{\sqrt{\Delta}}\left(\frac{\omega}{A}
+\frac12\right).\label{e1e2}
\end{equation}
Analogously obtain
\begin{align*}
E_1+E_2&=2\imath (X_{2Å"}+X_{31})-\imath (X_{32}+X_{23})
-\imath (c_1^{(3)}+c_1^{(2)})\\
&=\imath( V^{(3)} - V^{(2)} )(2 \left(\mathrm{ln}\, \vartheta_1(r) \right)' -
\left(\mathrm{ln}\, \vartheta_1(2r) \right)')  -\frac{\imath
  \lambda_2\lambda_3}{2\Delta}\\
&=-\frac{\imath\lambda_2\lambda_3}{\Delta}\left(\frac{\omega}{A}
+\frac12\right).
\end{align*}
To complete computation of $E_{1,2}$ we must find relation between
period $A$ of the curve $X$ and period $2\omega$ of the curve
$\widetilde{X}$. We shall do that by substituting (\ref{elsol})
to the (\ref{abc}) which should lead to equivalence. We have

\begin{align*}
&q_{1x}q_2-q_{2x}q_1=-4\delta_2\delta_3\mathrm{exp}[ (E_1+E_2)x ]\\
&\times \left\{\frac{V}{\vartheta_1(Vx')^2}
(\vartheta_1(Vx'-r)\vartheta_1'(Vx'+r) -
\vartheta_1(Vx'+r)\vartheta_1'(Vx'-r)   ) \right.\\
&\left. \qquad\qquad+ \frac{E_2-E_1}{\vartheta_1(Vx')^2} \vartheta_1(Vx'-r)
\vartheta_1(Vx'+r)
     \right\},
\end{align*}
where $x'=x-\frac{1}{2V}(1+\tau)  $. Substituting instead of
derivatives $\vartheta_1'(z)$ Weierstrass $\zeta$-functions we
transform the expression in the curly brackets to
\begin{align*}
&\frac{\vartheta_1(Vx'-r)\vartheta_1(Vx'+r)}{\vartheta_1(Vx')^2}
\left[ 2\omega V \zeta(2Vx'\omega+2\omega r)  \right.\\
&-\left. 2\omega V\zeta(2Vx'\omega-2\omega r)-8V\omega\eta
  r+E_2-E_1\right]\\
&=\sigma(2\omega
r)^2\mathrm{exp}\left\{-\frac{r^2\eta}{\omega}\right\}\wp(2\omega V
x')\\
&\times
\left[4\omega V\zeta(2\omega r)-8V\eta\omega r+E_2-E_1
+\frac{2\omega\sqrt{g_3} }{\wp(2\omega V x' )}\right],
 \end{align*}
where we applied addition formula for the Weierstrass $\sigma$ and
$\zeta$-functions
and took into account (\ref{wpzero}). Because this quantity should be
a constant with respect to $x$ we set
\begin{equation}E_1-E_2=4\omega V(\zeta(2\omega r) -2\eta
r), \quad E_1+E_2=0\label{e1e2a} \end{equation}
what in combination with (\ref{e1e2}) gives relation (\ref{aomega}) between
$\mathfrak{a}$-periods of the curve $X$ and $\widetilde{X}$. Therefore
the derivation of expressions for $E_{1,2}$ given in (\ref{en})
is completed. But let us continue the computation of the integral
level. We have now
\[\vert q_{1x}q_2-q_{2x}q_1\vert^2
=256\left|\omega^4\delta_2^2\delta_3^2g_3
\frac{\vartheta_1(r)^2}{\vartheta_1(0)^2}\right|.\]
But
\[ \delta_2^3\delta_3^2=\frac{\imath\mu_0}{256 A^4\Delta}
\frac{\vartheta_1(0)^2}{\vartheta_1(r)^2}  \quad
\text{and}\quad g_3=-64\Delta,  \]
what completes the derivation of equivalence.

It remains to show that $N_{1,2}=0$. Develop expression for $N_1$. It
is the sum of 3 terms which are
\begin{align*}
&-c_2^{(2)}-c_2^{(1)}=-\frac{\eta\omega(-3\lambda_2^2+10\imath
 \mu_0+\lambda_2\sqrt{\Delta})}{8A^2 \Delta}-\frac{\imath\mu_0\lambda_3^2
 (\imath\mu_0-\lambda_2^2+\lambda_2\sqrt{\Delta})
 }{\Delta^2(\lambda_2-\sqrt{\Delta})^2}\\
&2(W^{(1)}- W^{(2)}+W^{(3)})(\mathrm{ln}\,\vartheta_1(r) )'-
2W^{(3)}(\mathrm{ln}\,\vartheta_1(2r)
 )'=\frac{\imath\lambda_3^2\mu_0\omega}{A\Delta^2}\\
&2((V^{(1)})^2+(V^{(2)})^2+(V^{(3)})^2)(\mathrm{ln}\,\vartheta_1(r) )''-
2(V^{(3)})^2(\mathrm{ln}\,\vartheta_1(2r) )''\\
&=\frac{\eta\omega(-3\lambda_2^2+10\imath
 \mu_0+\lambda_2\sqrt{\Delta})}{8A^2 \Delta}
-\frac{\omega^2\lambda_3^2(\lambda_2+\sqrt{\Delta})^2}{4A^2\Delta^2}.
\end{align*}
Taking into the account equality (\ref{wpzero}) and substituting then
(\ref{aomega}) to the sum we obtain necessary
equality. The equality $N_2=0$ is derived in analogous way.

It could be also of interest to check by direct substitution that
solution (\ref{elsol}) satisfies to the Manakov system. To check that
we first compute

\begin{align}\label{term1}
&2(|q_1(x,t)|^2+|q_2(x,t)|^2)=8(\delta_2^2+\delta_3^2)
\frac{\vartheta_3(Vx'+r)\vartheta_3(Vx'-r)}{\vartheta_3(Vx)^2}\\
&=-\frac{\vartheta_1'(0)}{8\omega^2}\frac{\vartheta_3(Vx'+r)\vartheta_3(Vx'-r)}
{\vartheta_3(Vx)^2\vartheta_1(r)^2}-\frac{\sigma(2\omega
V x +2\omega r)\sigma(2\omega V x -2\omega r)  }{2
  \sigma^2(2*\omega V)\sigma^2(2\omega r) }\notag\\
&=-\frac12\wp(2\omega V x),\notag
\end{align}
where we used again in the Weierstrass addition theorem
(\ref{wpzero}). Further the first derivative

\begin{align*} \frac{\partial }{\partial x} &q_1(x,t)\\&=q_1(x,t)\left(
2V\omega\zeta(2\omega r) -4V\eta\omega-E+\frac{V\omega(\wp'(2\omega V)
  -\imath\sqrt{g_3})}{\wp(2\omega V x )}
\right) \end{align*} The first 3 terms in brackets vanish because of
expression for $E$ given in  (\ref{en}). Therefore

\begin{align} &\frac{\partial^2 }{\partial x^2} q_1(x,t) \label{term2}  \\
&=q_1(x,t)
\left[\left(\frac{V\omega(\wp'(2\omega V)
  -\imath\sqrt{g_3})}{\wp(2\omega V x )}
\right)^2+\frac{\partial}{\partial x}  \left(\frac{V\omega(\wp'(2\omega V)
  -\imath\sqrt{g_3})}{\wp(2\omega V x )}
\right)     \right]\notag\\
&=\frac12 q_1(x,t)\wp(2\omega V x).\notag
\end{align}
Combining (\ref{term1}) and (\ref{term2}) we obtain the equiality
(\ref{manakov1}). Validity of the (\ref{manakov2}) is proved
analogously.

\section*{Acknowledgements}The authors are grateful to
J.C.Eilbeck with the help in making the plots. They are also thank
to E.Previato for the pointing of the paper \cite{jo92}. VZE is
grateful to ESPRC for support under grant No GR/R2336/01 and to the
Issac Newton Institute for support  within the``In\-teg\-rable
systems'' programme in 2001 when the work on the paper was started
on. Informatics and Modelling of the Technical University of Denmark
and the MIDIT center are also acknowledged for funding of his
research visit in October-November 2003 within Grant 21-02-0500 from
the Danish Natural Science Research Council when this paper was
completed. ARI was supported in part by NSF Grant DMS-0099812 and by
Imperial College of the University of London via the EPSRC Grant.

\newcommand{\etalchar}[1]{$^{#1}$}
\providecommand{\bysame}{\leavevmode\hbox to3em{\hrulefill}\thinspace}


\end{document}